\documentclass[acmtog]{acmart}
\acmSubmissionID{1155}

\usepackage{booktabs} %

\usepackage[ruled]{algorithm2e} %

\SetAlFnt{\small}
\SetAlCapFnt{\small}
\SetAlCapNameFnt{\small}
\SetAlCapHSkip{0pt}

\usepackage[export]{adjustbox}
\usepackage{subcaption}

\copyrightyear{2024}
\acmYear{2024}
\setcopyright{acmlicensed}\acmConference[SA Conference Papers '24]{SIGGRAPH Asia 2024 Conference Papers}{December 3--6, 2024}{Tokyo, Japan}
\acmBooktitle{SIGGRAPH Asia 2024 Conference Papers (SA Conference Papers '24), December 3--6, 2024, Tokyo, Japan}
\acmDOI{10.1145/3680528.3687692}
\acmISBN{979-8-4007-1131-2/24/12}

\begin{document}
\title{Synchronize Dual Hands for Physics-Based Dexterous Guitar Playing}

\author{Pei Xu}
\orcid{0000-0001-7851-3971}
\affiliation{%
 \institution{Stanford University}
 \country{USA}}
\email{peixu@stanford.edu}
\author{Ruocheng Wang}
\orcid{0009-0005-8404-6405}
\affiliation{%
 \institution{Stanford University}
 \country{USA}}
\email{rcwang@cs.stanford.edu}

\begin{abstract}
We present a novel approach to synthesize dexterous motions for physically simulated hands in tasks that require coordination between the control of two hands with high temporal precision.
Instead of directly learning a joint policy to control two hands,
our approach performs bimanual control through cooperative learning where each hand is treated as an individual agent.
The individual policies for each hand are first trained separately, 
and then synchronized through latent space manipulation in a centralized environment to serve as a joint policy for two-hand control.
By doing so, we avoid directly performing policy learning in the joint state-action space of two hands with higher dimensions, greatly improving the overall training efficiency.
We demonstrate the effectiveness of our proposed approach in the challenging guitar-playing task.
The virtual guitarist trained by our approach can synthesize motions from unstructured reference data
of general guitar-playing practice motions, and accurately play diverse rhythms with complex chord pressing and string picking patterns based on the input guitar tabs that do not exist in the references.
Along with this paper, we provide the motion capture data that we collected as the reference for policy training.
Code is available at: \url{https://pei-xu.github.io/guitar}.
\end{abstract}



\begin{teaserfigure}
\centering
    \includegraphics[width=\linewidth]{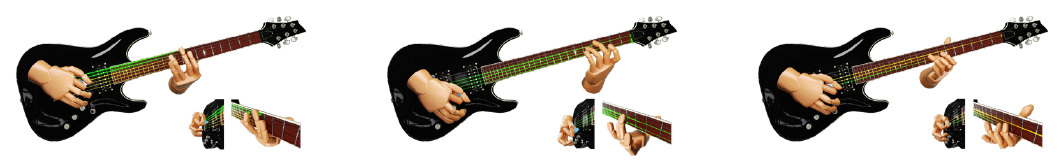}
    \caption{Examples demonstrating our approach to physics-based bimanual control for guitar playing with synchronized dual-hand motions.
    Yellow strings indicate the ones correctly pressed by the left hand and ready for the right hand to play, while green denotes those already played correctly.
    Policies trained using our approach can effectively coordinate the behaviors of two hands, and perform songs across diverse genres, tempos and playing techniques.
    From left to right, the performed songs are: \textit{Snow (Hey Oh)} by \textit{Red Hot Chili Peppers}, \textit{Summer} by \textit{Joe Hisaishi} and \textit{Spring Festival Overture} by \textit{Huanzhi Li}.}
 \label{fig:teaser}
\end{teaserfigure}

\maketitle

\section{Introduction}\label{sec:intro}
Facilitated by the advancements in computer graphics and VR/AR techniques, virtual music performance has emerged in recent years to provide a better immersive experience of live music~\cite{skreinig2022guitARhero}.
However, due to the high requirement of control precision, it is a challenge to effectively control virtually simulated characters to interact with musical instruments in a human-like and physically plausible way.
Solutions relying on motion capture recordings to reproduce motions virtually, though guaranteeing the realism of the animation, are typically costly and time-consuming, and usually unable to synthesize motions for unseen songs or rhythms.  
In this paper,
we present an approach to synthesize motions for physically animated hands and fingers performing novel music virtually, specially, using the instrument of guitar.

The guitar is a popular instrument in daily life. 
However, mastering the guitar is a difficult task which needs a long time of practice,
as it requires dexterity and coordination in both hands to fret chords (pressing multiple strings on the guitar neck), strum strings and execute intricate fingerpicking patterns, etc.
During guitar performance, two hands undertake heterogeneous tasks interacting with strings but have to coordinate to perform temporally synchronized behaviors with exceptional spatial precision.
To produce desired sounds,
the left hand needs to fret (press at or release from) strings rapidly in accordance with the music rhythm, 
while the right hand must pluck the strings
when they are in the correct pressed or released states.
Similarly, some playing techniques, e.g. natural harmonics, require the left hand to wait for the right hand and release a pressed string right after it is played.
Previous work explored approaches to string fretting~\cite{heijink2002complexity, radisavljevic2004path, wortman2021CombinoChordAG} relying on heuristic rules to find out finger placements,
but ignores the dynamics of real human hands in the temporal context during guitar playing. 
Our work focuses on data-driven approaches to generate human-like fretting motions that are physically plausible, exhibiting the grace and dexterity of human guitarists. 
Further, to provide a complete guitar-playing experience, we introduce bimanual control to generate motions where the fretting and strumming hands play a given song on a virtual guitar at the same time 
with sounds generated through post-processing based on the two hands' dynamics.

In computer graphics, previous works of hand motion synthesis mostly focus on object grasping and/or manipulating ~\cite{andrychowicz2020learning,zhang2021manipnet,yang2022learning,xie2023hierarchical,zhao2013robust}, or hand gesture generation~\cite{liang2022seeg,qian2021speech,Zhang2024SemanticGesture,kucherenko2020gesticulator}.
Effective bimanual control with synchronized but heterogeneous tasks, like guitar playing, is under-investigated. 
Typically, bimanual control can be performed by considering two hands as a unified agent.
However, training one joint policy for two hands would exponentially increase the dimension of the state-action space, compared with the case of having only one hand.
Meanwhile, controlling two hands in a centralized environment would introduce more computation costs for simulation.

Instead of regarding two hands as a unified agent and performing policy training jointly,
we consider guitar playing as a cooperative task where the left and right hands are treated as individual agents.
To improve the learning efficiency,
our approach decouples the two-hand control first by training the policies for individual hands separately in decentralized environments,
and then quickly adapts single-hand policies to generate synchronized behaviors in a centralized environment by manipulating the latent spaces of the two individual policies.
Our approach can
improve the overall learning efficiency by reducing the training required for cooperative learning in centralized environments.

While string fretting and playing are goal-driven tasks during guitar playing,
we perform physics-based control with additional imitation learning to ensure the naturalness and physical plausibility of the generated motions in such contact-rich tasks.
We captured a set of hand motions approximately one hour long from a professional guitarist.
All the motions are general guitar-playing techniques rather than the tracking of playing any specific songs.
While taking two short clips of scaling and strumming motions as the references for imitation learning,
our approach shows excellent performance in synthesizing those unstructured reference motions to generate natural guitar-playing motions with automatic fingering inference for novel string fretting and playing patterns.

We evaluate our approach
both qualitatively and quantitatively.
As shown in the experiment section,
while learning high-quality single-hand control,
our approach can efficiently synchronize single-hand policies and
effectively coordinate two hands' behaviors to provide dexterous motions with high precision both temporally and spatially.
Additional ablation studies in the supplementary materials support the proposed training scheme for bimanual control along with highlighting the potential of our approach to apply to tasks other than guitar playing.

\section{Related Works}
Traditional methods for physics-based control typically rely on trajectory optimization and/or human-designed heuristic rules for control~\cite{liu2008synthesis,liu2009dextrous,mordatch2012contact,wang2013video,ye2012synthesis,chen2023synthesizing}.
Early exploration of imitation learning using pre-collected mocap data~\cite{pollard2005physically,kry2006interaction,zhao2013robust} achieved impressive results to generate human-like motions.
In recent years, with the advancement of deep learning techniques,
reinforcement learning based imitation learning~\cite{deepmimic,amp,ase,scadiver, motionvae,controlvae,yao2023moconvq,gail2,iccgan} has drawn lots of attention and become a popular approach for physics-based character control.
Our approach follows the literature and focuses on physics-based dexterous bimanual control using reinforcement learning.

\subsection{Dexterous Control for Physically Simulated Hands}
Dexterous control has attracted significant attention due to its complexity and frequent occurrence in everyday situations.
Typical dexterous control tasks can generally be divided into two main categories: object grasping and/or relocation~\cite{xie2023hierarchical,zhao2013robust,liu2009dextrous} and in-hand manipulation~\cite{andrychowicz2020learning,zhang2021manipnet,yang2022learning}.
The common scenarios of these tasks do not have a high requirement of temporal precision,
and often only need one hand.
In contrast, our guitar-playing task requires precise control of two hands simultaneously.

Generating human-like instrument-playing motions based on given music has been studied extensively in recent years 
~\cite{Liu2020BodyMG, shlizerman2018audio, li2018skeleton, kao2020temporally,chen2021guzheng,hirata2022audio}.  Early work~\cite{elkoura2003handrix} considers motion generation for guitar playing.
However, they only focus on the left hand's fretting motions rather than the coordinated motions of two hands.
Besides, their method is kinematics-based, ignoring the constraints of finger transitioning under the temporal context and the complicated collision and contact states between fingers and those between the fretboard and fingers.

Our goal in this paper is to generate physically plausible two-hand motions for guitar playing.
Instead of using heuristic rules to perform control and motion generation for virtual musical instrument playing~\cite{zhu2013piano,elkoura2003handrix,zakka2023robopianist},
we employ imitation learning plus reinforcement learning to perform motion synthesis from unstructured reference data~\cite{composite}, accounting for the dexterity and grace of human guitarists, and rely on the control policy to perform fingering inference through exploration on its own without needing to explicitly design any heuristic rules for motion generation.

\subsection{Hand Motion Datasets}
Currently, most hand motion datasets only provide motions involving object grasping or manipulation~\cite{taheri2020grab,chao2021dexycb,fan2023arctic}.
Only a few datasets contain motions of playing musical instruments like violin and piano~\cite{simon2017hand, Grauman2023EgoExo4DUS}.
To our knowledge, there is still a lack of publicly available datasets containing high-quality motions of guitar playing.
Along with this paper,
we provide the guitar-playing motions that we captured from a professional guitarist.
The whole dataset is around 1 hour long, including the practice of various guitar-playing techniques.
From the dataset, we take two clips of scaling and strumming motions for our policy training.
We refer to the supplementary materials for the details of our dataset and the motion capture setup.

\section{Overview}
An overview of our proposed policy synchronization system for learning physics-based bimanual control is shown in Fig.~\ref{fig:overview}.
To avoid directly performing training in a centralized environment with two hands,
in this system,
we treat the left and right hands as independent agents, with their control policies trained separately.
The
two single-hand control policies are
finally
put into a centralized environment for cooperative learning to generate synchronized two-hand motions.
Instead of directly fine-tuning the two pre-trained single-hand policies,
we introduce a synchronizer to quickly coordinate two policies' behaviors via 
manipulating the latent spaces $\mathbf{z}_t^L$ and $\mathbf{z}_t^R$ for the left and right hand respectively.

For single-hand policy training,
our approach takes the multi-objective reinforcement learning framework~\cite{composite}, considering both the motion imitation and task objectives of fret pressing for the left hand and string picking for the right hand.
The single-hand policies are trained for general purposes using music notes extracted from real songs and those generated procedurally for the left and right hands respectively.
For policy synchronization, we extend the policy adaptation strategy from~\cite{adaptnet} and introduce a synchronizer to coordinate two control policies simultaneously
through latent space manipulation to generate guitar-playing motions for specific target songs.
To provide a better understanding of the whole system, 
we will first elaborate on the synchronization process in Section~\ref{sec:sync}, and then detail the implementation of policy training for the dexterous dual-hand guitar playing task in Section~\ref{sec:guitar_playing}. 

\section{Policy Synchronization}
\label{sec:sync}
Under the setup of reinforcement learning,
A two-agent cooperative task can generally be considered an optimization problem to maximize the cumulative rewards associated with two agents jointly, i.e.,
    $\max \mathbb{E}_{\tau\sim p(\tau)}\left[\sum_t \gamma^{t-1} \mathcal{R}_t\right]$
where $\tau=\{\mathbf{o}_t^L, \mathbf{o}_t^R, \mathbf{a}_t^L, \mathbf{a}_t^R\}_t$ is the state-action trajectory of the two agents denoted by $L$ and $R$ respectively, $\gamma$ is a discount factor, $\mathcal{R}_t$ is a scalar reward to evaluate the two agents' joint performance at the time step $t$. 
A joint action policy $\pi(\mathbf{a}_t^L, \mathbf{a}_t^R | \mathbf{o}_t^L, \mathbf{o}_t^R)$ can be achieved in a centralized environment by the same way that we train single-agent policies.

\begin{figure}[t]
    \centering
    \includegraphics[width=\linewidth]{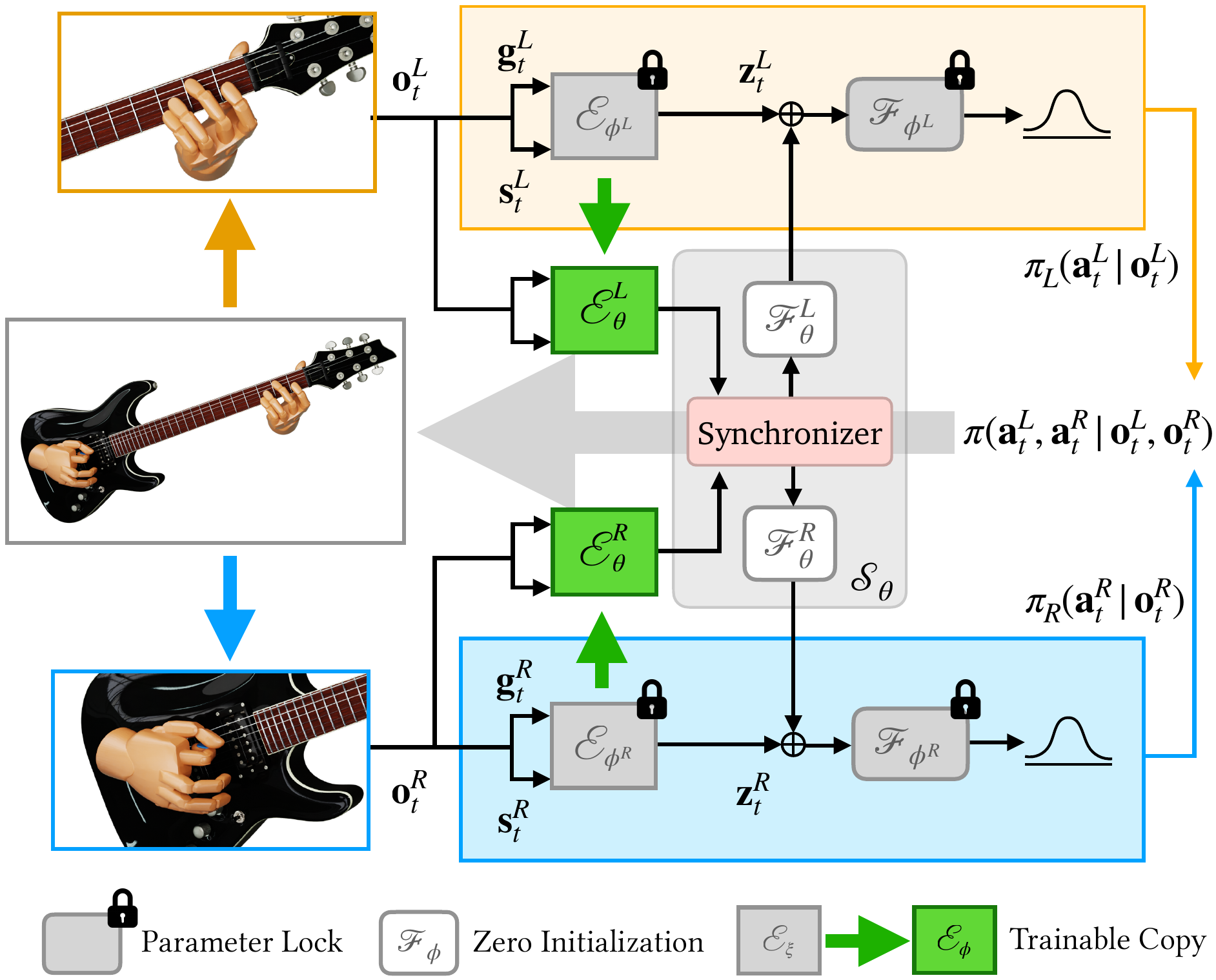}
    \caption{Overview of the proposed system synchronizing dual policies for dexterous guitar playing with two hands. Our system performs two-hand policy training in two steps. First, we decentralize the control of two hands, and train the left-hand policy for fret pressing (orange box) and the right-hand policy for string picking (blue box) independently in a decentralized manner. Then, we lock the previously trained single-hand policies, and introduce a synchronizer to coordinate the behaviors of single-hand policies in a centralized training environment to obtain a joint policy for two-hand control.
    The synchronization is achieved quickly by modifying single-hand policies' behavior patterns through latent space manipulation.}
    \label{fig:overview}
\end{figure}

Inspired by how humans perform cooperative tasks,
we propose a two-step training scheme to perform two-agent cooperative learning efficiently.
Our scheme first decouples the cooperative task into a set of sub-tasks for individual agents and performs policy training independently for each agent, which is like that humans learn basic work skills before being involved in a cooperative task;
and then the two individual policies, $\pi_L(\mathbf{a}_t^L | \mathbf{o}_t^L)$ and $\pi_R(\mathbf{a}_t^R | \mathbf{o}_t^R)$, are adapted in a centralized environment to serve as a joint policy $\pi(\mathbf{a}_t^L, \mathbf{a}_t^R | \mathbf{o}_t^L, \mathbf{o}_t^R)$ providing synchronized cooperative behaviors for two-agent control.
Through this setting, 
we avoid directly performing optimization in the joint state-action space and, meanwhile, improve the overall training efficiency by preventing directly performing cooperative learning for two agents lacking basic skills for cooperation.

As we assume that the individual agents are already trained well during the first individual training phase and have the potential to provide needed behaviors for cooperation,
our goal during policy synchronization is to quickly coordinate individual policies' behavior patterns conditionally for cooperation rather than letting them learn new behaviors from scratch.
Therefore,
we do not directly fine-tune the networks of the pre-trained individual policies.
Instead, as shown in Fig.~\ref{fig:overview}
we introduce a synchronizer to take the input from both individual policies
and conditionally coordinate their behaviors by outputting offsets to the two policies' latents.
During synchronization,
all parameters in the pre-trained policy networks are locked, and
we only optimize the synchronizer-related parameters $\theta$.
Though we show the same network structure for individual policies, $\pi_L(\mathbf{a}_t^L | \mathbf{o}_t^L)$ and $\pi_R(\mathbf{a}_t^R | \mathbf{o}_t^R)$, in Fig.~\ref{fig:overview},
there is no requirement for them to be identical in practice.

Following the previous literature~\cite{adaptnet},
the synchronizer in our implementation is built using a multilayer perceptron~(MLP) architecture but with the last fully-connected layers initialized by zero ($\mathcal{F}_\phi^L$ and $\mathcal{F}_\phi^R$ in Fig.~\ref{fig:overview}).
This means that the synchronizer will not change the behavior of individual policies when synchronization training starts.
The policy exploration in the synchronization phase, therefore, will start not randomly but from the trajectory generated by the pre-trained individual policies,
and thus would be more efficient.
Additionally,
we take advantage of the pre-trained individual policies,
and initialize the synchronizer's state encoders, $\mathcal{E}_\theta^L$ and $\mathcal{E}_\theta^R$, by duplicating parameters from the corresponding state encoders in the pre-trained policies ($\mathcal{E}_{\phi^L}$ and $\mathcal{E}_{\phi^R}$ respectively in Fig.~\ref{fig:overview}).
The process of latent space manipulation can be written as:
\begin{equation}
    \mathbf{z}_t^L = \mathcal{E}_{\phi^L}(\mathbf{o}_t^L) + \Delta \mathbf{z}_t^L, \text{and } \mathbf{z}_t^R = \mathcal{E}_{\phi^R}(\mathbf{o}_t^R) + \Delta \mathbf{z}_t^R
\end{equation}
where $\mathbf{z}_t^L$ and $\mathbf{z}_t^R$ are the latents
of the two policies $\pi_L$ and $\pi_R$, and
\begin{equation}
    \Delta \mathbf{z}_t^L, \Delta \mathbf{z}_t^R = \mathcal{S}_\theta\left(\textsc{Concat}\left(\mathcal{E}_\theta^L(\mathbf{o}_t^L), \mathcal{E}_\theta^R(\mathbf{o}_t^R)\right) \right).
\end{equation}
During the synchronization phase,
the joint policy can be updated by optimizing the synchronizer with parameter $\theta$ through general reinforcement learning algorithms.

\section{Let's Play Guitar}\label{sec:guitar_playing}
In Fig.~\ref{fig:guitar},
we show the profile of a common guitar model.
During guitar playing, typically,
the left hand moves along the fretboard to press one or more target strings at the positions of different string-fret combinations, 
while the right hand interacts with the strings using fingers and/or through a plectrum called \textit{pick}.
In the following, we use the terminology \textit{chords} to describe cases where more than one string needs to be pressed by the left hand at the same time; and for the right hand, we focus only on the case where a pick held between the thumb and index is used to interact with the strings.

\subsection{Left-Hand Policy for Fret Pressing}
\begin{figure}[t]
    \centering
    \includegraphics[width=\linewidth]{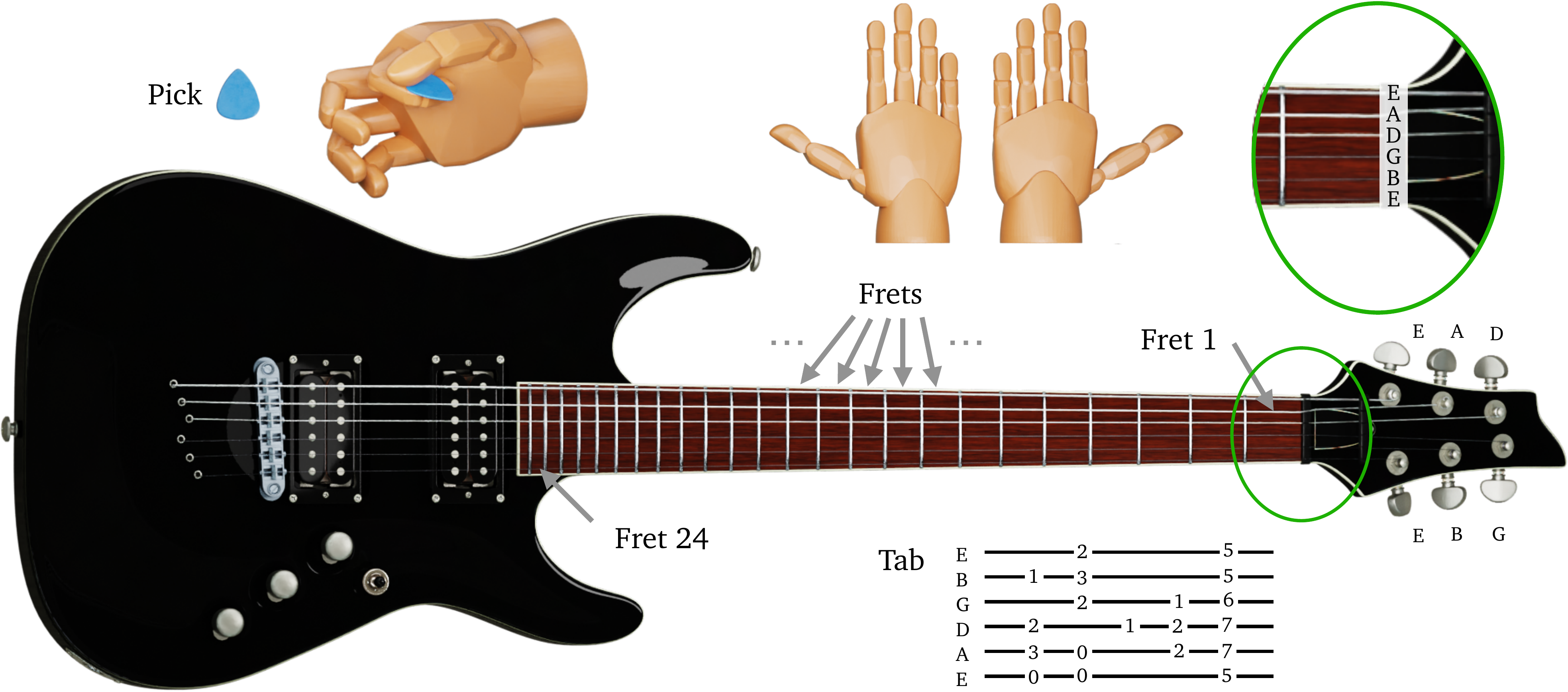}
    \caption{Profile of the simulated guitar and hand models. Our guitar model is left-handed. It has six strings with twenty-four frets and is played with a pick. Given a common guitar chord spanning at most four frets plus two additional conditions -- open string (being picked without fret pressing) and mute string (not being picked to generate any sound) -- for each string, there are nearly 900,000 chord combinations theoretically, which makes it difficult to be fully mastered. Each of our hand models has 16 links with 27 degrees of freedom (DoFs), where the wrist joint has 6 DoFs, the MCP joints have 2 DoFs with the exception that the thumb MCP having 3, and all the other finger joints have 1 DoF. The simulated guitar has a scale length of 24.5 inches (around 0.62m) in our implementation. In this work, we only consider the control of hands and assume that the guitar is fixed in the 3D space. We can obtain right-handed guitar-playing motions by mirroring the setup of the hand and guitar.}
    \label{fig:guitar}
\end{figure}

Guitar tablature (tab), as shown in Fig.~\ref{fig:guitar}, is a common visual representation of music notes for guitar playing, which indicates the target strings for picking and the target fret at which a string should be pressed before being picked while 0 means an open string that should be picked without any fret being pressed.
We take the format of tab as the input goal state to the left-hand policy, but using -1 to represent open strings and 0 to represent mute strings.
We consider a horizon of five future notes, and along with each note, a timer variable is employed to indicate the remaining duration of that note in terms of the number of simulation frames.
This results in a goal state vector $\mathbf{g}_t^L \in \mathbb{R}^{5\times7}$.
The pose state $\mathbf{s}_t$ is composed of the position, orientation, and linear and angular velocities of each link in the hand model related to the guitar.
We take a 2-frame historical observation, which gives us a state vector $\mathbf{s}_t \in \mathbb{R}^{2\times16\times13}$.

Fingering strategies for fret pressing involve the chord patterns of the current and future notes and depend on the biological structure (movement constraints, energy consumption, etc.) of human hands.
Besides using fingertips to press single strings, some chords need one finger to press multiple strings or need multiple fingers to press different strings at the same fret simultaneously.
Those challenges make the fingering problem intricate. 
In our implementation,
we do not explicitly provide additional fingering inference, but rely on the control policy itself to explore fret-pressing strategies.

Since every string has its own target pressing condition at each time step $t$,
we consider the fret-pressing task as a multi-objective problem,
where the target pressing condition of each string is regarded as an objective.
Given the three possible conditions, we define the goal-driven reward function for each string $i$ as
\begin{equation}\label{eq:l}%
    r_{i} = \begin{cases}
        r_{i}^{k} & \text{if string $i$ should be pressed at fret $k$ for picking,}  \\
        r_{i}^{0} & \text{if string $i$ should not be pressed (open string),}  \\
        r_{i}^{-1} & \text{otherwise (mute string).}  \\
    \end{cases}
\end{equation}

To encourage fret pressing, $r_{i}^{k}$ is defined based on the \textit{minimal} distance from the target fret-string position $\mathbf{p}_{i,k}$ and all \textit{available} finger parts at the time step $t$:
\begin{equation}\label{eq:r_ik}
    r_{i}^{k} = 0.8\exp(-1000 d_{i,k}^2) + 0.2 \exp(-30 d_{i,k}^2)
\end{equation}
where $d_{i,k} = \min_j ||\mathbf{p}_{j\rightarrow i,k} - \mathbf{p}_{i,k}||$, $\mathbf{p}_{i,k}$ is the intersection position of string $i$ and the middle line of fret $k$, and $\mathbf{p}_{j\rightarrow i,k}$ is the closest point on an \textit{available} finger part $j$ to $\mathbf{p}_{i,k}$. Each finger is considered to have three cylinder-shaped parts between its MCP, PIP, DIP, and tip. The hand model has 12 finger parts totally, which can be used for string fretting, excluding the thumb. The closest point $\mathbf{p}_{j\rightarrow i,k}$ can be found by calculating the distance between the point $\mathbf{p}_{i,k}$ to the axis segment of the finger-part cylinder $j$.

\begin{figure}[t]
    \centering
    \includegraphics[width=\linewidth]{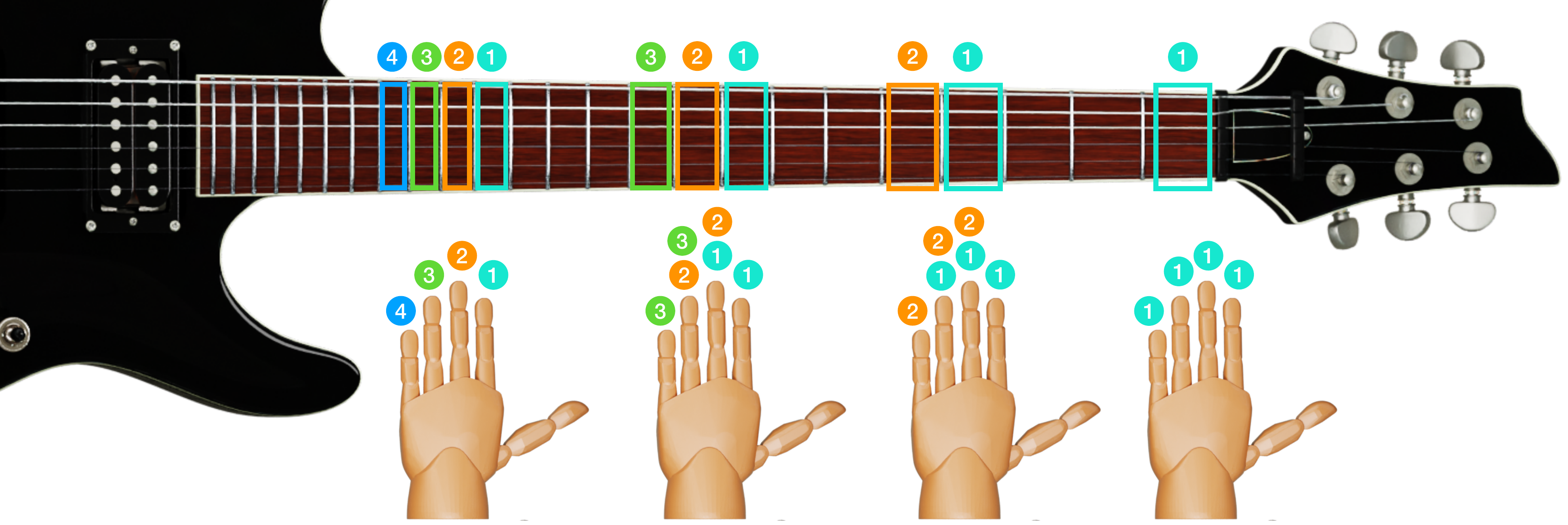}
    \caption{Demonstrations of finger availability for fret pressing. We allow multiple fingers to press at one fret but do not allow fingers to cross over. Each finger can press at no more than one fret, and at most four frets could be pressed simultaneously. Here we show the four possible cases with the one (right) to four (left) target frets that need to be pressed. Target frets and the corresponding available fingers that can be used to press at the target frets are labeled using the same number.}
    \label{fig:finger}
\end{figure}

In practice, when using a finger part other than tips to press a string,
the finger typically has to lie down on the fret to perform a strong pressing.
Hence, we consider
a finger part between MCP and PIP or that between PIP and DIP \textit{available} for string pressing only if the angle between it and the fretboard surface is less than $5^\circ$.
In challenging cases where strings at
multiple frets need to be pressed simultaneously,
one finger typically would not be used to perform pressing across multiple frets or cross over other fingers in order to generate a clear sound.
Therefore, we additionally decide if a finger's three parts are \textit{available} for pressing at strings on a target fret based on the discrepancy between the finger order and the order of the target fret, as demonstrated in Fig.~\ref{fig:finger}.
We compute $r_i^k$ only considering \textit{available} finger parts.

Since the fretboard is around 0.45m long with fret widths varying from 0.01m to 0.035m,
$r_{i}^k$ in Eq.~\ref{eq:r_ik} is designed as the sum of two exponential functions, where the former one with a higher weight ensures that the function is sensitive enough when $d_{i,k}$ is around 0.01m, while the later one prevents the function falling into the saturation range when $d_{i,k}$ is near 0.45m.

For an open string that should be picked without being pressed, $r_{i}^{0}$ in Eq.~\ref{eq:l} induces fingers not to touch the string. It is defined based on the minimal distance between the string and all fingers: 
\begin{equation}
    r_{i}^{0} = \min\left\{(d_{i}/0.007)^2, 1\right\}
\end{equation}
where the $d_{i}$ is the shortest distance between string $i$ and all 12 finger parts.
The constant scalar $0.007$ is chosen since our hand model has fingers with a radius of around 0.006m. 
Therefore, $d_{i}/0.007$ indeed is a normalized distance, and guarantees that string $i$ is not touched by any finger, when it is equal to or greater than 1.

The third case in Eq.~\ref{eq:l} is for a mute string, 
which is assumed not to generate any sound during the period of a note.
As mute strings will not be picked, their pressing conditions are not strictly constrained.
However, to produce clear sounds, mute strings theoretically should be touched slightly (not pressed) to mute their vibration if they were picked and still vibrating.
Here, we use a loose constraint
\begin{equation}
    r_{i}^{-1} = 0.9+0.1r_{i}^0
\end{equation}
to encourage but not force fingers to keep away from the string.

To take into account the left hand's overall performance,
two common terms are added to $r_{i}$ for all the six string-based goal-driven objectives. The final reward for each objective $i$ is
\begin{equation}\label{eq:r_il}
    r_{t,i}^L = 0.8r_{i} + 0.2r_\text{correct}^L - 0.05r_\text{energy}^L
\end{equation}
where $r_\text{correct}^L = 1$ if all strings are pressed at the target frets correctly or $0$ otherwise, and $r_\text{energy}^L$ is a term measuring the energy consumption based on the linear velocities of wrist and fingers:
\begin{equation}\label{eq:r_energy_L}
    r_\text{energy}^L = \exp\left(-\left(||\mathbf{v}_w|| + 0.1\sum\nolimits_f ||\mathbf{v}_f||\right)^2\right)
\end{equation}
where $\mathbf{v}_f$ is the linear velocity of each fingertip locally related to the wrist, and $\mathbf{v}_w$ is the velocity of the wrist in the global space.

To generate natural motions,
we also employ two imitation objectives.
Following the previous literature of composite motion learning~\cite{composite},
we decouple the left-hand motion into (1) the wrist-thumb motion related to the guitar and (2) the local motions of the other four fingers related to the wrist,
and use a GAN-like architecture~\cite{iccgan} to perform imitation learning.
Such a decoupling strategy ensures that the hand can naturally hold the guitar through a correct wrist and thumb pose provided by the reference motions,
while the pose of the other four fingers is free from the guitar's pose.

Written in a general on-policy learning form, the final optimization objective for policy training is:
\begin{equation}\label{eq:lh_policy_optimization}
    \max \mathbb{E}_t\left[\sum\nolimits_\kappa w_\kappa \bar{A}_{t,\kappa}^L \log \pi_L(\mathbf{a}_t^L | \mathbf{o}_t^L) \left| \phi^L \right. \right]
\end{equation}
where $\bar{A}_{t,\kappa}^L$ is the standardized advantage that is estimated according to the goal-driven or imitation objective reward $r_{t,\kappa}^L$, and $w_\kappa$ is the weight associated with each object $\kappa$. 
We have $w_\kappa=0.15$ for each of the six goal-driven objectives of fret pressing and $w_\kappa=0.05$ for each of the two imitation objectives.
Following the previous literature, we use DPPO~\cite{schulman2017proximal} as the backbone reinforcement learning algorithm which takes an actor-critic architecture for policy optimization.

\subsection{Right-Hand Policy for String Picking}\label{sec:right}
The right-hand policy performs string picking and thus only needs to know if a string should be picked or not.
Therefore, compared to the left-hand policy,
the goal state $\mathbf{g}_t^R$ is defined using only a binary vector for tab representation without the fretting information.
During the duration of one note,
a target string is expected to be picked only once.
To let the policy ignore strings that are already picked,
if a target string is picked,
its corresponding representation in $\mathbf{g}_t^R$ will be set to 0 in the remaining frames of that note.
While the left hand's performance of fretting at each time step $t$ is decided only by the fretting state at that time step, 
the right hand's performance of picking is evaluated by the picking behavior during the duration of each note.
To provide better state-value estimation,
the critic network additionally takes a binary vector indicating if each string is tackled wrongly throughout the current note.
This leads to an extended goal state $\mathbf{g}_t^{R+} \in \mathbb{R}^{5\times7+6}$ for the critic network.
Such an implementation is crucial for efficient policy training.

To generate sounds in unison, when multiple strings need to be picked,
they must be picked in rapid succession with a consistent order from either up to down or down to up.
Therefore, differing from the left-hand policy that has multiple string-based goal-driven objectives,
we consider the string-picking task as an integrated objective.
The reward function is designed to evaluate the picking performance at the time step $t$ or the overall performance during the duration of the current note if no target string is left:
\begin{equation}\label{eq:r}%
    r_\text{pick} = \begin{cases}
        r^{+} & \text{if any string is picked at the time step $t$,} \\
        r^{-} & \text{if no target string is left for picking,} \\
        r^{\times} & \text{otherwise.}
    \end{cases}
\end{equation}

$r^{+}$ is to evaluate the picking behavior if it happens. Since we require strings to be picked in a consistent order, a string is considered tackled wrongly (1) if the string is a non-target string but picked, (2) if the string is not picked but a string below it is picked when the picking direction is top to down, or (3) if the string is not picked but a string above it is picked when the picking direction is down to up.
We define $r^{+}$ based on the distance from the pick tip to the wrongly tackled string(s) and encourage the pick to move away to the farthest wrongly tackled strings, i.e.
\begin{equation}%
    r^{+} = \begin{cases}
    \min_i r_{d_i} & \text{if any string $i$ is tackled wrongly at step $t$,} \\
    r_\text{succ}  & \text{otherwise,} \\
    \end{cases}
\end{equation}
where $r_{d_i}$ is a reward defined by the shortest distance from the pick tip to the wrongly tackled string $i$.
Here we employ a form similar to Eq.~\ref{eq:r_ik} to define $r_{d_i}$ as
\begin{equation}
    r_{d_i} = 0.175 \exp(-10000 d_i^2) + 0.025 \exp(-2000 d_i^2)
\end{equation}
taking into account that the distance between the pick tip to a string during guitar playing would vary from millimeters to decimeters.
$r^{+} = r_\text{succ} = 1$ if not any string is tackled wrongly, namely, the top-most string(s) are picked from up to down or the bottom-most string(s) are picked down to up.

$r^{-}$ in Eq.~\ref{eq:r} is used to evaluate the overall picking performance during the current note. It takes effect only when all target strings have been picked or if there is not any target string in the current note for picking. $r^{-}$ is defined as
\begin{equation}
    r^{-} = 0.4 \min\left\{\min\nolimits_i d_i/0.003, 1\right\} + 0.1 \sum\nolimits_i r_i + 0.5\prod\nolimits_i r_i
\end{equation}
where $i = 1, \cdots, 6$ indicates a string, $d_i$ is the shortest distance from the pick tip to the string $i$, $r_i = 1$ for a target string if it has been picked correctly or for a non-target string if it was not picked accidentally, or $0$ otherwise.
The term $\prod_i r_i$ is an additional reward effecting only when all six strings are tackled correctly.
Considering the string radius, here we expect the pick tip to be at least 0.003m away from any strings.

We take $r^\times$ to entice correct picking behaviors if one or more strings need to be picked but no picking happens at the time step $t$:
\begin{equation}
    r^\times = 0.2 + 2\min\{r_{d_\text{top}}, r_{d_\text{bottom}}\}
\end{equation}
where the subscript \textit{top} and \textit{bottom} indicate the top and bottom-most target strings respectively.
$r^\times$ encourages the policy to move the pick to the nearest top or bottom-most target string.
At each time step $t$, we always have $1 = r_\text{succ} > r^\times \geq 0.2 \geq \min_i r_{d_i}$, which brings a higher priority for the policy to move to and pick strings from the top or bottom-most target one.
    
Besides the picking performance, we consider the contact condition of the thumb and index fingers for pick holding and the energy consumption caused by the hand and pick movement. The overall goal-driven reward for the right-hand policy training is:
\begin{equation}\label{eq:r_ir}
    r_t^R = r_\text{pick} + 0.05 r_\text{contact} + 0.05 (r_\text{energy}^R + r_\text{energy}^\text{pick})
\end{equation}
where
$r_\text{contact} = 1$ if the thumb and index contact with each other correctly to hold the pick or $0$ otherwise; $r_\text{energy}^R$ is the reward for the energy consumption of the hand's movement and defined in the same way with $r_\text{energy}^L$ in Eq.~\ref{eq:r_energy_L}; and $r_\text{energy}^\text{pick}$ is a term used to produce stable movement of the pick, and is defined based on the pick's linear velocity $\mathbf{v}_\text{pick}$ and acceleration $\mathbf{a}_\text{pick}$:
\begin{equation}
    r_\text{energy}^\text{pick} = \exp(-20||\mathbf{v}_\text{pick}||^2) - 14||0.05 \mathbf{a}_\text{pick}||^4.
\end{equation}

Similar to the left-hand policy training,
we rely on imitation learning to generate natural human-like motions.
Nevertheless, the right-hand motion using a pick to strum strings, though not as diverse as the left-hand motion, has more subtle dynamics.
As the hand sweeps up and down, fingers would show compliant motions with the movement of the wrist.
Therefore, we consider the whole right hand as a group for motion imitation.
The right-hand policy is optimized using the formula of Eq.~\ref{eq:lh_policy_optimization} but with only two objectives (one goal-driven objective $r_t^R$ and one imitation objective) having an equal weight of $0.5$.

\subsection{Dual Hand Policy Synchronization}\label{sec:guitar_sync}
After achieving two single-hand policies,
we perform policy synchronization using the approach depicted in Section~\ref{sec:sync}.
We adopt the same multi-objective learning method~(cf. Eq.~\ref{eq:lh_policy_optimization}) to optimize the joint policy for specific target songs during synchronization, but update only the synchronizer-related parameter $\theta$ rather than the pre-trained policy networks.
We keep the six goal-driven objectives and two imitation objectives for the left hand.
For the right hand, however, we take only the goal-driven objective of string picking without any imitation learning,
as we found that the right-hand policy can effectively master picking motions 
without needing further training given the limited number of string-picking patterns, while the left hand still needs the reference motion to help learn natural motion poses when facing novel chord patterns.
To introduce the additional cooperative condition between two hands, 
when computing the reward $r_\text{pick}$ for the right hand,
it will be regarded as there is no target (using the $r^{\times}$ case in Eq.~\ref{eq:r}) unless all strings are at the expected pressing states.

\section{Experiments}
We run experiments using IsaacGym~\cite{makoviychuk2021isaac} as the physical engine and taking a hand model modified from the Modular Prosthetic Limb model~\cite {kumar2015mujoco}.
Our hand model, as shown in Fig.~\ref{fig:guitar}, is skeleton-driven where each joint is modeled as a motor controlled through a PD controller, which results in an action space $\mathbf{a}_t^L, \mathbf{a}_t^R \in \mathbb{R}^{27}$.
All simulation runs at 240Hz with the control policies running at 60Hz.
Due to the complication of simulating string dynamics,
all strings are modeled virtually rather than physically simulated.
A string is considered pressed by a finger part if the distance between them is less than 0.006m (approximately the radius of fingers).
A string is assumed to be picked if the pick tip goes through that string's virtual position from one side to the other side with a trajectory below the string.

We perform general-purpose training for the left-hand policy using notes extracted from 500 music tracks with a wide range of rhythm styles and tempos.
We further augment the extracted notes online during training by randomly shifting them along the fretboard, while keeping the chord pattern unchanged, plus a stochastic tempo adjustment of $\pm 20$ beats per minute (BPM).
For the right hand,
we use procedurally generated strumming patterns as the target goals for training.
During synchronization,
behaviors of the two pre-trained individual policies are fine-tuned using specific target songs.
We take (1) a clip of scaling practice motions around 40 seconds long including two hands and (2) a clip of the right hand's strumming motion which is about 3 seconds long
as the references to synthesize motions for novel music note playing instead of performing motion tracking.
We refer to the supplementary materials for additional implementation and data collection details.

\subsection{Numeric Results}
\paragraph{Metrics:} 
We use F1 scores to evaluate the trained policies' performance on the level of music notes.
A true positive sample in our case means a string that is correctly pressed or picked during the duration of a note.
Precision indicates the ratio of true positive samples to the total number of strings that are pressed or picked.
Recall indicates the percentage of the correctly handled strings among the target strings of one note.
For the left hand, a string is considered pressed correctly if it is pressed at the target fret continuously for at least 2/3 of the duration of the note.
For the right hand, a string is considered picked correctly if it is a target string and is picked in a consistent order
(cf. Section~\ref{sec:right}) once and only once during the note's duration.
For the joint policy of two hands, a string is treated as being handled correctly only if (1) the string is picked correctly, (2) the string is being pressed correctly when being picked, and (3) the pressing state should be kept for at least 2/3 duration of the note after the string is picked.

\begin{figure}[t]
    \centering
    \includegraphics[width=\linewidth]{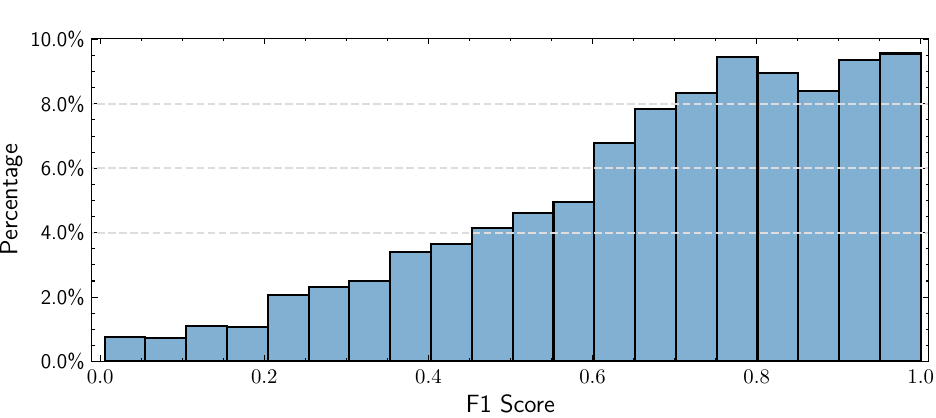}
    \caption{Distribution of F1 scores achieved by the left-hand policy when playing chords. The test set contains 50 music tracks with 1721 unrepeated measures having 4859 chords.}
    \label{fig:f1_chords}
\end{figure}

\paragraph{Single-hand policy evaluation:}
We evaluate the performance of the separately trained single-hand policies using a test set containing 50 music tracks.
We remove all the repeated measures and keep 4859 notes left.
The overall F1 score achieved by the left-hand policy is around 0.8.
For regular notes where only one string needs to be pressed,
the policy obtains almost 100\% accuracy even for fast songs with a tempo as high as 150BPM, which is challenging for human players.
For chord notes playing, %
the left-hand policy can obtain an F1 score above 0.8 for around 40\% of the tested chords.
The distribution of the F1 scores for chords playing is shown in Fig.~\ref{fig:f1_chords}.
From the figure, we can see that
there are around 20\% chords for which the policy achieves an F1 score of less than 0.5.
We find that the poorly performed chords are those not in or only appearing few times in the training set for the left hand.
Though we employ a large set of 500 music tracks for the left-hand policy training,
it is not enough to fully cover all the possible chords that may appear during guitar playing.
In contrast,
the right hand policy only could face a limited number of picking patterns, and achieves consistent performance with an average F1 score above 0.9 for all the tested notes.
Given the sophistication of rhythm and chord generation,
it would be an interesting direction for future work to design programs that can provide diverse chords with a better balanced distribution for the left-hand policy to learn.

\begin{figure}[t]
    \centering
    \includegraphics[width=\linewidth]{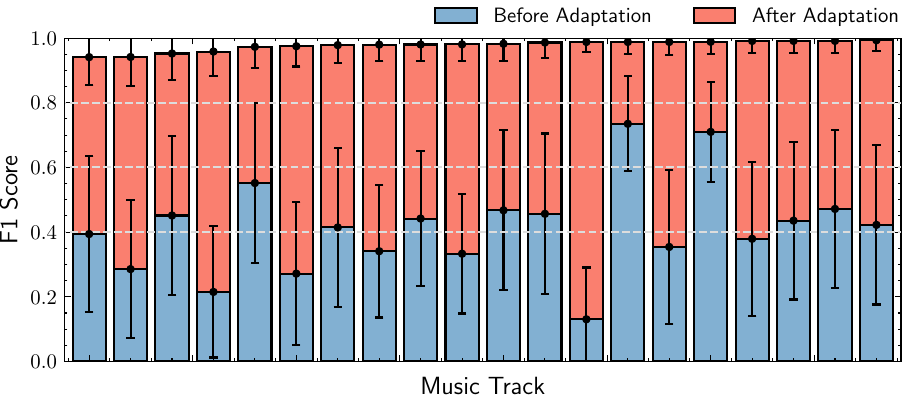}
    \caption{F1 scores of bimanual performance on a test set of 25 music tracks before and after synchronization. We show both the mean values and the standard deviations of F1 scores evaluated on the level of music notes.}
    \label{fig:f1_sync}
\end{figure}

\paragraph{Synchronization performance evaluation:}
To evaluate the effectiveness of the proposed policy synchronization scheme,
we perform policy synchronization for 25 music tracks using the pre-trained single-hand policies,
and report the F1 scores before and after synchronization in Fig.~\ref{fig:f1_sync}.
Due to the cooperation requirement,
the F1 scores involving two hands typically are lower compared to the case when evaluating the single-hand policies independently.
As shown in the figure, policy synchronization brings about a significant improvement in F1 scores, some of which are as high as 657\% with an average improvement of 193\%.
An interesting observation is that we only perform latent space manipulation to change the policies' behavior patterns rather than fine-tuning the whole policy networks, and the right-hand policy even does not perform any imitation learning during synchronization.
However, the single-hand policies' performance still could be improved effectively. 
This means that the single-hand policies were trained well during the pre-training phase to %
provide poses potentially needed during guitar playing.
Overall, the two-hand policy can achieve almost 100\% accuracy for most tested songs after synchronization.
Such improvement partially is because of the fine-tuning effect during policy synchronization,
but also comes from the synchronized two-hand behaviors.
To further demonstrate that,
we studied the change in F1 scores when some nearly perfect single-hand policies were taken for synchronization.
We refer to the supplementary materials for the related analysis.

\begin{figure*}[t]
    \includegraphics[width=\linewidth]{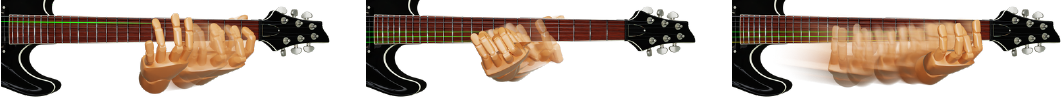}
    \caption{Examples of the left-hand policy swiftly and accurately pressing strings at the target frets in the distance. Strings highlighted in green are target strings that are pressed correctly. Policies trained by our approach can perform accurate long-distance pressing using multiple fingers at the same time.}
    \label{fig:fret}
\end{figure*}

\begin{figure*}
    
    
    
    \includegraphics[width=.98\linewidth]{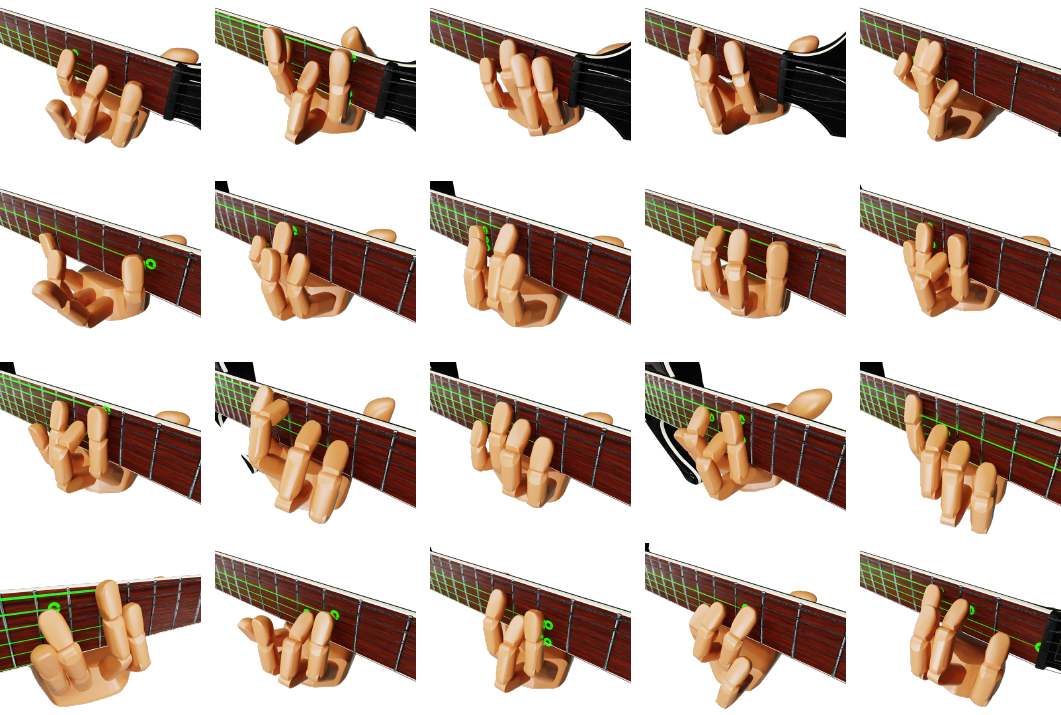}
    \caption{Examples of the diverse chords performed by the left-hand policy. Green spots indicate the target string-fret positions while some of them are occluded by the hand due to accurate fretting.}
    \label{fig:chords}
\end{figure*}

\begin{figure*}[t]
    \includegraphics[width=.98\linewidth]{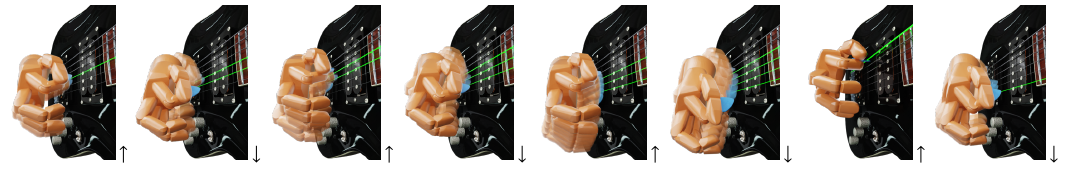}
    \caption{Complex picking patterns during the right-hand policy playing the song \textit{Hotel California}. While accurately picking the target strings, our policy follows an efficient trajectory to move up and down and up again consistently. The arrows indicate the picking direction.}
    \label{fig:pick}
\end{figure*}

\begin{figure*}[t]
    \includegraphics[width=.98\linewidth]{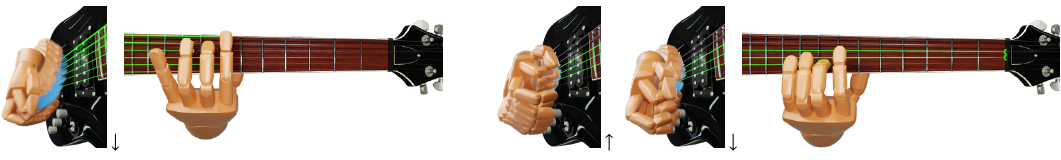}
    \caption{Fast string strumming to play chords. Correctly strummed target strings are shown in green. Our policy can perform strumming smoothly in rapid succession with high accuracy. The right figure shows a pair of consecutive strumming, where strings are picked from down to up and then up to down.}
    \label{fig:strum}
\end{figure*}

\begin{figure*}[t]
\begin{minipage}[b]{0.28\linewidth}
  \centering
  
\includegraphics[width=.98\linewidth]{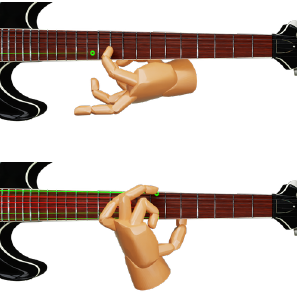}
  \caption{Motions generated by policies trained using only goal-directed task rewards without motion imitation.}
  \label{fig:noimit}
\end{minipage}
\hfill
\begin{minipage}[b]{0.66\linewidth}

    \includegraphics[width=.98\linewidth]{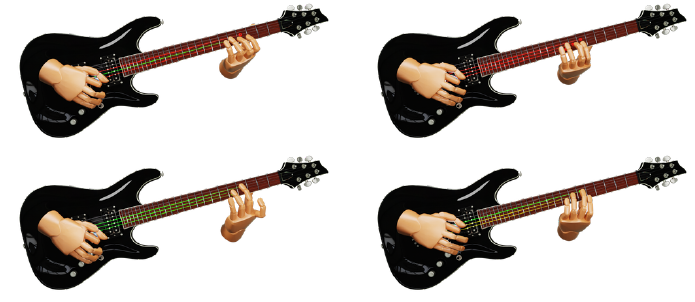}
    \caption{Two-hand performance before (top) and after (bottom) policy synchronization. Red strings indicate the target strings that are picked without being correctly fretted. Yellow strings in the right-bottom figure indicate those that are correctly fretted and have not been picked.}
  \label{fig:sync}
\end{minipage}
\end{figure*}

\begin{figure}\centering
    
    
    
    \includegraphics[width=.95\linewidth]{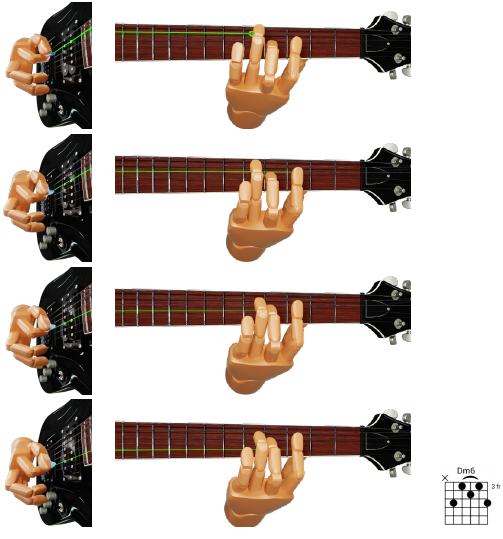}
    \caption{Demonstrations of arpeggios. While the right hand picks individual strings following the target input notes (from top to bottom), the left hand keeps a stable pose for chord pressing by taking the chord as its target goal.}
    \label{fig:arpeggios}
\end{figure}

\subsection{Qualitative Evaluation}
We show the snapshots of our control policies' live performance in Fig.~\ref{fig:teaser}.
Policies trained by our approach can be synchronized for music with distinct tempos, genres and playing techniques, and generate visually convincing animations for novel rhythms not included in the reference motion data.

For the left hand, we highlight the control policy's accurate fretting ability in Fig.~\ref{fig:fret}.
As can be seen, the hand can perform long-distance movements swiftly to fret target strings.
The wrist part's traveling distance in the 3rd case, for example, crosses 11 frets and reaches around 0.3m. 
Note that the maximal fret is 0.035m wide while the minimal is just around 0.01m. Such small widths request a demand for precise control spatially.
In Fig.~\ref{fig:chords},
we show the dexterous control of the left hand performing 
diverse novel chords. %
All chords shown in the figure do not exist in the reference motions used for policy training.
We also include some challenging chord-pressing tasks. For example, the 1st one in the 2nd row requires a long stretch between fingers; the 2nd one in the 1st row and the 3rd in the 2nd row require one finger to press multiple strings on the same fret; and the 1st one in the 4th row require different fingers to press strings on one fret (the ring finger on the 2nd and 3rd strings and pinkie on the 6th).
In Fig.~\ref{fig:noimit}, we exhibit the results when the left-hand policy is trained without motion imitation.
By comparing those results, we can see that our approach can effectively synthesize the limited poses in the reference data and generate human-like dexterous motions.

For the right hand, maintaining a steady rhythm during picking and strumming is a crucial skill for guitar playing. 
While Fig.~\ref{fig:pick} shows the picking performance of the right hand going up and down rhythmically, Fig.~\ref{fig:strum} shows the demonstrations of the right hand performing strumming where multiple strings are picked rapidly for successive notes. 
Without explicitly indicating picking directions, the right-hand policy can perform picking energy efficiently and maintain a regular up-to-down and down-to-up rhythm.
In Fig.~\ref{fig:arpeggios},
we exhibit the policy's performance of maintaining consistent rhythm while playing arpeggios, where the strings of one chord are picked individually by the right hand and the left hand keeps a chord-pressing pose.
In this demonstration, we explicitly feed a \textit{Dm6} chord (shown at the right-bottom corner in the figure) as the goal state input to the left-hand policy only, while the synchronizer and right-hand policy just take the picked string as the goal.
Such an implementation is the same as what we could do in real life to put a chord indicator along the tab as guidance for fretting.

In Fig.~\ref{fig:sync}, we compare the policy's performance before and after synchronization.
The highly synchronized behaviors with exceptional precision in timing enable our approach for music ensembles using multiple guitars.
We refer to the supplementary video for the animated results of ensemble playing.

\section{Conclusions}
We present a novel approach for physics-based dual hand control in tasks requiring cooperation between two hands.
Our approach first decouples the control of two hands in decentralized environments and then synchronizes the pre-trained single-hand policies' behaviors for cooperation in a centralized environment.
We demonstrate the proficiency of our approach in challenging tasks of guitar playing that require dexterous bimanual control with high precision both temporally and spatially.
We refer to the supplementary materials for ablation studies and discussions about the limitations of the proposed method.
Overall, our approach allows effective motion synthesis from unstructured references, providing high-quality guitar-playing motions in terms of both accuracy and naturalness for novel songs with diverse genres and tempos.
Additionally, we captured a set of diverse guitar-playing motions from a professional guitarist.
All the data is released publicly available for future work.

\begin{figure}
    \includegraphics[width=\linewidth]{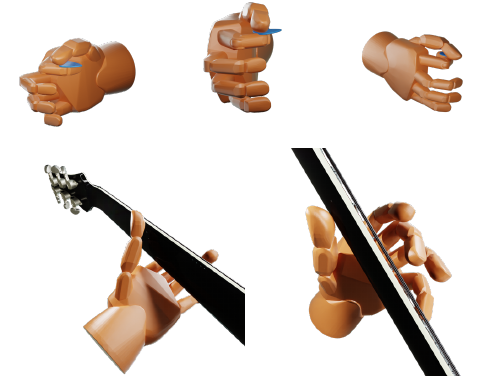}
    \caption{Poses of the right hand holding the pick stably (top) and poses of the left hand contacting the guitar fretboard in a natural way.}
    \label{fig:hold}
\end{figure}

\begin{acks}
This work was supported in part by Stanford Institute for Human-Centered Artificial Intelligence.
The motion capture solutions and devices were provided by NOKOV\footnote{https://www.nokov.com}.
The authors would like to thank Ryan Canales from Clemson University for proposing the initial idea to perform guitar playing and for providing the initial data to verify the proposed method, and Prof.~Ioannis Karamouzas
from UC Riverside and
Prof.~Sophie J\"oerg from University of Bamberg for the meaningful discussions.
We would also like to thank Hao Wang, the subject guitarist of our motion capture, for his performance,
Mengyao Zhang and Dr. Hongming Fan from UIUC for their help in organizing motion capturing,
and Dr. Shuchun Sun and Jichao Zhao from Clemson University for their help in developing the motion capture pipeline.
\end{acks}

\bibliographystyle{ACM-Reference-Format}
\bibliography{sample-bibliography}


\begin{thebibliography}{52}


\ifx \showCODEN    \undefined \def \showCODEN     #1{\unskip}     \fi
\ifx \showDOI      \undefined \def \showDOI       #1{#1}\fi
\ifx \showISBNx    \undefined \def \showISBNx     #1{\unskip}     \fi
\ifx \showISBNxiii \undefined \def \showISBNxiii  #1{\unskip}     \fi
\ifx \showISSN     \undefined \def \showISSN      #1{\unskip}     \fi
\ifx \showLCCN     \undefined \def \showLCCN      #1{\unskip}     \fi
\ifx \shownote     \undefined \def \shownote      #1{#1}          \fi
\ifx \showarticletitle \undefined \def \showarticletitle #1{#1}   \fi
\ifx \showURL      \undefined \def \showURL       {\relax}        \fi
\providecommand\bibfield[2]{#2}
\providecommand\bibinfo[2]{#2}
\providecommand\natexlab[1]{#1}
\providecommand\showeprint[2][]{arXiv:#2}

\bibitem[Andrychowicz et~al\mbox{.}(2020)]%
        {andrychowicz2020learning}
\bibfield{author}{\bibinfo{person}{OpenAI:~Marcin Andrychowicz}, \bibinfo{person}{Bowen Baker}, \bibinfo{person}{Maciek Chociej}, \bibinfo{person}{Rafal Jozefowicz}, \bibinfo{person}{Bob McGrew}, \bibinfo{person}{Jakub Pachocki}, \bibinfo{person}{Arthur Petron}, \bibinfo{person}{Matthias Plappert}, \bibinfo{person}{Glenn Powell}, \bibinfo{person}{Alex Ray}, {et~al\mbox{.}}} \bibinfo{year}{2020}\natexlab{}.
\newblock \showarticletitle{Learning dexterous in-hand manipulation}.
\newblock \bibinfo{journal}{\emph{The International Journal of Robotics Research}} \bibinfo{volume}{39}, \bibinfo{number}{1} (\bibinfo{year}{2020}), \bibinfo{pages}{3--20}.
\newblock


\bibitem[Chao et~al\mbox{.}(2021)]%
        {chao2021dexycb}
\bibfield{author}{\bibinfo{person}{Yu-Wei Chao}, \bibinfo{person}{Wei Yang}, \bibinfo{person}{Yu Xiang}, \bibinfo{person}{Pavlo Molchanov}, \bibinfo{person}{Ankur Handa}, \bibinfo{person}{Jonathan Tremblay}, \bibinfo{person}{Yashraj~S Narang}, \bibinfo{person}{Karl Van~Wyk}, \bibinfo{person}{Umar Iqbal}, \bibinfo{person}{Stan Birchfield}, {et~al\mbox{.}}} \bibinfo{year}{2021}\natexlab{}.
\newblock \showarticletitle{DexYCB: A benchmark for capturing hand grasping of objects}. In \bibinfo{booktitle}{\emph{Conference on Computer Vision and Pattern Recognition}}.
\newblock


\bibitem[Chen et~al\mbox{.}(2023a)]%
        {chen2021guzheng}
\bibfield{author}{\bibinfo{person}{Jiali Chen}, \bibinfo{person}{Changjie Fan}, \bibinfo{person}{Zhimeng Zhang}, \bibinfo{person}{Gongzheng Li}, \bibinfo{person}{Zeng Zhao}, \bibinfo{person}{Zhigang Deng}, {and} \bibinfo{person}{Yu Ding}.} \bibinfo{year}{2023}\natexlab{a}.
\newblock \showarticletitle{A Music-Driven Deep Generative Adversarial Model for Guzheng Playing Animation}.
\newblock \bibinfo{journal}{\emph{IEEE Transactions on Visualization and Computer Graphics}} \bibinfo{volume}{29}, \bibinfo{number}{2} (\bibinfo{year}{2023}), \bibinfo{pages}{1400--1414}.
\newblock


\bibitem[Chen et~al\mbox{.}(2023b)]%
        {chen2023synthesizing}
\bibfield{author}{\bibinfo{person}{Sirui Chen}, \bibinfo{person}{Albert Wu}, {and} \bibinfo{person}{C~Karen Liu}.} \bibinfo{year}{2023}\natexlab{b}.
\newblock \showarticletitle{Synthesizing Dexterous Nonprehensile Pregrasp for Ungraspable Objects}. In \bibinfo{booktitle}{\emph{ACM SIGGRAPH 2023 Conference Proceedings}}. \bibinfo{pages}{1--10}.
\newblock


\bibitem[Chung et~al\mbox{.}(2014)]%
        {chung2014empirical}
\bibfield{author}{\bibinfo{person}{Junyoung Chung}, \bibinfo{person}{Caglar Gulcehre}, \bibinfo{person}{Kyunghyun Cho}, {and} \bibinfo{person}{Yoshua Bengio}.} \bibinfo{year}{2014}\natexlab{}.
\newblock \showarticletitle{Empirical Evaluation of Gated Recurrent Neural Networks on Sequence Modeling}. In \bibinfo{booktitle}{\emph{NIPS 2014 Workshop on Deep Learning}}.
\newblock


\bibitem[ElKoura and Singh(2003)]%
        {elkoura2003handrix}
\bibfield{author}{\bibinfo{person}{George ElKoura} {and} \bibinfo{person}{Karan Singh}.} \bibinfo{year}{2003}\natexlab{}.
\newblock \showarticletitle{Handrix: Animating the Human Hand}.
\newblock  (\bibinfo{year}{2003}), \bibinfo{pages}{110–119}.
\newblock


\bibitem[Fan et~al\mbox{.}(2023)]%
        {fan2023arctic}
\bibfield{author}{\bibinfo{person}{Zicong Fan}, \bibinfo{person}{Omid Taheri}, \bibinfo{person}{Dimitrios Tzionas}, \bibinfo{person}{Muhammed Kocabas}, \bibinfo{person}{Manuel Kaufmann}, \bibinfo{person}{Michael~J Black}, {and} \bibinfo{person}{Otmar Hilliges}.} \bibinfo{year}{2023}\natexlab{}.
\newblock \showarticletitle{ARCTIC: A dataset for dexterous bimanual hand-object manipulation}. In \bibinfo{booktitle}{\emph{Conference on Computer Vision and Pattern Recognition}}. \bibinfo{pages}{12943--12954}.
\newblock


\bibitem[Grauman et~al\mbox{.}(2024)]%
        {Grauman2023EgoExo4DUS}
\bibfield{author}{\bibinfo{person}{Kristen Grauman}, \bibinfo{person}{Andrew Westbury}, \bibinfo{person}{Lorenzo Torresani}, \bibinfo{person}{Kris Kitani}, \bibinfo{person}{Jitendra Malik}, \bibinfo{person}{Triantafyllos Afouras}, \bibinfo{person}{Kumar Ashutosh}, \bibinfo{person}{Vijay Baiyya}, \bibinfo{person}{Siddhant Bansal}, \bibinfo{person}{Bikram Boote}, \bibinfo{person}{Eugene Byrne}, \bibinfo{person}{Zachary Chavis}, \bibinfo{person}{Joya Chen}, \bibinfo{person}{Feng Cheng}, \bibinfo{person}{Fu-Jen Chu}, \bibinfo{person}{Sean Crane}, \bibinfo{person}{Avijit Dasgupta}, \bibinfo{person}{Jing Dong}, \bibinfo{person}{Mar{\'i}a Escobar}, \bibinfo{person}{Cristhian Forigua}, \bibinfo{person}{Abrham~Kahsay Gebreselasie}, \bibinfo{person}{Sanjay Haresh}, \bibinfo{person}{Jing Huang}, \bibinfo{person}{Md~Mohaiminul Islam}, \bibinfo{person}{Suyog~Dutt Jain}, \bibinfo{person}{Rawal Khirodkar}, \bibinfo{person}{Devansh Kukreja}, \bibinfo{person}{Kevin~J Liang}, \bibinfo{person}{Jia-Wei Liu},
  \bibinfo{person}{Sagnik Majumder}, \bibinfo{person}{Yongsen Mao}, \bibinfo{person}{Miguel Martin}, \bibinfo{person}{Effrosyni Mavroudi}, \bibinfo{person}{Tushar Nagarajan}, \bibinfo{person}{Francesco Ragusa}, \bibinfo{person}{Santhosh~K. Ramakrishnan}, \bibinfo{person}{Luigi Seminara}, \bibinfo{person}{Arjun Somayazulu}, \bibinfo{person}{Yale Song}, \bibinfo{person}{Shan Su}, \bibinfo{person}{Zihui Xue}, \bibinfo{person}{Edward Zhang}, \bibinfo{person}{Jinxu Zhang}, \bibinfo{person}{Angela Castillo}, \bibinfo{person}{Changan Chen}, \bibinfo{person}{Xinzhu Fu}, \bibinfo{person}{Ryosuke Furuta}, \bibinfo{person}{Cristina Gonzalez}, \bibinfo{person}{Prince Gupta}, \bibinfo{person}{Jiabo Hu}, \bibinfo{person}{Yifei Huang}, \bibinfo{person}{Yiming Huang}, \bibinfo{person}{Weslie Khoo}, \bibinfo{person}{Anush Kumar}, \bibinfo{person}{Robert Kuo}, \bibinfo{person}{Sach Lakhavani}, \bibinfo{person}{Miao Liu}, \bibinfo{person}{Mingjing Luo}, \bibinfo{person}{Zhengyi Luo}, \bibinfo{person}{Brighid Meredith},
  \bibinfo{person}{Austin Miller}, \bibinfo{person}{Oluwatumininu Oguntola}, \bibinfo{person}{Xiaqing Pan}, \bibinfo{person}{Penny Peng}, \bibinfo{person}{Shraman Pramanick}, \bibinfo{person}{Merey Ramazanova}, \bibinfo{person}{Fiona Ryan}, \bibinfo{person}{Wei Shan}, \bibinfo{person}{Kiran Somasundaram}, \bibinfo{person}{Chenan Song}, \bibinfo{person}{Audrey Southerland}, \bibinfo{person}{Masatoshi Tateno}, \bibinfo{person}{Huiyu Wang}, \bibinfo{person}{Yuchen Wang}, \bibinfo{person}{Takuma Yagi}, \bibinfo{person}{Mingfei Yan}, \bibinfo{person}{Xitong Yang}, \bibinfo{person}{Zecheng Yu}, \bibinfo{person}{Shengxin~Cindy Zha}, \bibinfo{person}{Chen Zhao}, \bibinfo{person}{Ziwei Zhao}, \bibinfo{person}{Zhifan Zhu}, \bibinfo{person}{Jeff Zhuo}, \bibinfo{person}{Pablo Arbel{\'a}ez}, \bibinfo{person}{Gedas Bertasius}, \bibinfo{person}{David~J. Crandall}, \bibinfo{person}{Dima Damen}, \bibinfo{person}{Jakob~Julian Engel}, \bibinfo{person}{Giovanni~Maria Farinella}, \bibinfo{person}{Antonino Furnari},
  \bibinfo{person}{Bernard Ghanem}, \bibinfo{person}{Judy Hoffman}, \bibinfo{person}{C.~V. Jawahar}, \bibinfo{person}{Richard~A. Newcombe}, \bibinfo{person}{Hyun~Soo Park}, \bibinfo{person}{James~M. Rehg}, \bibinfo{person}{Yoichi Sato}, \bibinfo{person}{Manolis Savva}, \bibinfo{person}{Jianbo Shi}, \bibinfo{person}{Mike~Zheng Shou}, {and} \bibinfo{person}{Michael Wray}.} \bibinfo{year}{2024}\natexlab{}.
\newblock \showarticletitle{Ego-Exo4D: Understanding Skilled Human Activity from First- and Third-Person Perspectives}. In \bibinfo{booktitle}{\emph{Conference on Computer Vision and Pattern Recognition}}.
\newblock


\bibitem[Heijink and Meulenbroek(2002)]%
        {heijink2002complexity}
\bibfield{author}{\bibinfo{person}{Hank Heijink} {and} \bibinfo{person}{Ruud~GJ Meulenbroek}.} \bibinfo{year}{2002}\natexlab{}.
\newblock \showarticletitle{On the complexity of classical guitar playing: functional adaptations to task constraints}.
\newblock \bibinfo{journal}{\emph{Journal of motor behavior}} \bibinfo{volume}{34}, \bibinfo{number}{4} (\bibinfo{year}{2002}), \bibinfo{pages}{339--351}.
\newblock


\bibitem[Hirata et~al\mbox{.}(2022)]%
        {hirata2022audio}
\bibfield{author}{\bibinfo{person}{Asuka Hirata}, \bibinfo{person}{Keitaro Tanaka}, \bibinfo{person}{Masatoshi Hamanaka}, {and} \bibinfo{person}{Shigeo Morishima}.} \bibinfo{year}{2022}\natexlab{}.
\newblock \showarticletitle{Audio-Driven Violin Performance Animation with Clear Fingering and Bowing}.
\newblock In \bibinfo{booktitle}{\emph{ACM SIGGRAPH 2022 Posters}}. \bibinfo{pages}{1--2}.
\newblock


\bibitem[Kao and Su(2020)]%
        {kao2020temporally}
\bibfield{author}{\bibinfo{person}{Hsuan-Kai Kao} {and} \bibinfo{person}{Li Su}.} \bibinfo{year}{2020}\natexlab{}.
\newblock \showarticletitle{Temporally guided music-to-body-movement generation}. In \bibinfo{booktitle}{\emph{Proceedings of the 28th ACM International Conference on Multimedia}}. \bibinfo{pages}{147--155}.
\newblock


\bibitem[Kingma(2014)]%
        {kingma2014adam}
\bibfield{author}{\bibinfo{person}{Diederik~P Kingma}.} \bibinfo{year}{2014}\natexlab{}.
\newblock \showarticletitle{Adam: A method for stochastic optimization}.
\newblock \bibinfo{journal}{\emph{arXiv preprint arXiv:1412.6980}} (\bibinfo{year}{2014}).
\newblock


\bibitem[Kry and Pai(2006)]%
        {kry2006interaction}
\bibfield{author}{\bibinfo{person}{Paul~G Kry} {and} \bibinfo{person}{Dinesh~K Pai}.} \bibinfo{year}{2006}\natexlab{}.
\newblock \showarticletitle{Interaction capture and synthesis}.
\newblock \bibinfo{journal}{\emph{ACM Transactions on Graphics (TOG)}} \bibinfo{volume}{25}, \bibinfo{number}{3} (\bibinfo{year}{2006}), \bibinfo{pages}{872--880}.
\newblock


\bibitem[Kucherenko et~al\mbox{.}(2020)]%
        {kucherenko2020gesticulator}
\bibfield{author}{\bibinfo{person}{Taras Kucherenko}, \bibinfo{person}{Patrik Jonell}, \bibinfo{person}{Sanne Van~Waveren}, \bibinfo{person}{Gustav~Eje Henter}, \bibinfo{person}{Simon Alexandersson}, \bibinfo{person}{Iolanda Leite}, {and} \bibinfo{person}{Hedvig Kjellstr{\"o}m}.} \bibinfo{year}{2020}\natexlab{}.
\newblock \showarticletitle{Gesticulator: A framework for semantically-aware speech-driven gesture generation}. In \bibinfo{booktitle}{\emph{Proceedings of the 2020 international conference on multimodal interaction}}. \bibinfo{pages}{242--250}.
\newblock


\bibitem[Kumar and Todorov(2015)]%
        {kumar2015mujoco}
\bibfield{author}{\bibinfo{person}{Vikash Kumar} {and} \bibinfo{person}{Emanuel Todorov}.} \bibinfo{year}{2015}\natexlab{}.
\newblock \showarticletitle{Mujoco haptix: A virtual reality system for hand manipulation}. In \bibinfo{booktitle}{\emph{2015 IEEE-RAS 15th International Conference on Humanoid Robots (Humanoids)}}. IEEE, \bibinfo{pages}{657--663}.
\newblock


\bibitem[Li et~al\mbox{.}(2018)]%
        {li2018skeleton}
\bibfield{author}{\bibinfo{person}{Bochen Li}, \bibinfo{person}{Akira Maezawa}, {and} \bibinfo{person}{Zhiyao Duan}.} \bibinfo{year}{2018}\natexlab{}.
\newblock \showarticletitle{Skeleton Plays Piano: Online Generation of Pianist Body Movements from MIDI Performance.}. In \bibinfo{booktitle}{\emph{ISMIR}}. \bibinfo{pages}{218--224}.
\newblock


\bibitem[Liang et~al\mbox{.}(2022)]%
        {liang2022seeg}
\bibfield{author}{\bibinfo{person}{Yuanzhi Liang}, \bibinfo{person}{Qianyu Feng}, \bibinfo{person}{Linchao Zhu}, \bibinfo{person}{Li Hu}, \bibinfo{person}{Pan Pan}, {and} \bibinfo{person}{Yi Yang}.} \bibinfo{year}{2022}\natexlab{}.
\newblock \showarticletitle{Seeg: Semantic energized co-speech gesture generation}. In \bibinfo{booktitle}{\emph{Proceedings of the IEEE/CVF Conference on Computer Vision and Pattern Recognition}}. \bibinfo{pages}{10473--10482}.
\newblock


\bibitem[Ling et~al\mbox{.}(2020)]%
        {motionvae}
\bibfield{author}{\bibinfo{person}{Hung~Yu Ling}, \bibinfo{person}{Fabio Zinno}, \bibinfo{person}{George Cheng}, {and} \bibinfo{person}{Michiel van~de Panne}.} \bibinfo{year}{2020}\natexlab{}.
\newblock \showarticletitle{Character controllers using motion VAEs}.
\newblock \bibinfo{journal}{\emph{ACM Trans. Graph.}} \bibinfo{volume}{39}, \bibinfo{number}{4}, Article \bibinfo{articleno}{40} (\bibinfo{year}{2020}).
\newblock


\bibitem[Liu(2008)]%
        {liu2008synthesis}
\bibfield{author}{\bibinfo{person}{C~Karen Liu}.} \bibinfo{year}{2008}\natexlab{}.
\newblock \showarticletitle{Synthesis of interactive hand manipulation}. In \bibinfo{booktitle}{\emph{Proceedings of the 2008 ACM SIGGRAPH/Eurographics Symposium on Computer Animation}}. \bibinfo{pages}{163--171}.
\newblock


\bibitem[Liu(2009)]%
        {liu2009dextrous}
\bibfield{author}{\bibinfo{person}{C~Karen Liu}.} \bibinfo{year}{2009}\natexlab{}.
\newblock \showarticletitle{Dextrous manipulation from a grasping pose}.
\newblock In \bibinfo{booktitle}{\emph{ACM SIGGRAPH 2009 papers}}. \bibinfo{pages}{1--6}.
\newblock


\bibitem[Liu et~al\mbox{.}(2020)]%
        {Liu2020BodyMG}
\bibfield{author}{\bibinfo{person}{Jun-Wei Liu}, \bibinfo{person}{Hung-Yi Lin}, \bibinfo{person}{Yu-Fen Huang}, \bibinfo{person}{Hsuan-Kai Kao}, {and} \bibinfo{person}{Li Su}.} \bibinfo{year}{2020}\natexlab{}.
\newblock \showarticletitle{Body Movement Generation for Expressive Violin Performance Applying Neural Networks}.
\newblock \bibinfo{journal}{\emph{International Conference on Acoustics, Speech and Signal Processing (ICASSP)}} (\bibinfo{year}{2020}), \bibinfo{pages}{3787--3791}.
\newblock


\bibitem[Makoviychuk et~al\mbox{.}(2021)]%
        {makoviychuk2021isaac}
\bibfield{author}{\bibinfo{person}{Viktor Makoviychuk}, \bibinfo{person}{Lukasz Wawrzyniak}, \bibinfo{person}{Yunrong Guo}, \bibinfo{person}{Michelle Lu}, \bibinfo{person}{Kier Storey}, \bibinfo{person}{Miles Macklin}, \bibinfo{person}{David Hoeller}, \bibinfo{person}{Nikita Rudin}, \bibinfo{person}{Arthur Allshire}, \bibinfo{person}{Ankur Handa}, {and} \bibinfo{person}{Gavriel State}.} \bibinfo{year}{2021}\natexlab{}.
\newblock \bibinfo{title}{Isaac Gym: High Performance GPU-Based Physics Simulation For Robot Learning}.
\newblock
\newblock
\showeprint[arxiv]{2108.10470}~[cs.RO]


\bibitem[Merel et~al\mbox{.}(2017)]%
        {gail2}
\bibfield{author}{\bibinfo{person}{Josh Merel}, \bibinfo{person}{Yuval Tassa}, \bibinfo{person}{Dhruva TB}, \bibinfo{person}{Sriram Srinivasan}, \bibinfo{person}{Jay Lemmon}, \bibinfo{person}{Ziyu Wang}, \bibinfo{person}{Greg Wayne}, {and} \bibinfo{person}{Nicolas Heess}.} \bibinfo{year}{2017}\natexlab{}.
\newblock \showarticletitle{Learning human behaviors from motion capture by adversarial imitation}.
\newblock \bibinfo{journal}{\emph{arXiv preprint arXiv:1707.02201}} (\bibinfo{year}{2017}).
\newblock


\bibitem[Mordatch et~al\mbox{.}(2012)]%
        {mordatch2012contact}
\bibfield{author}{\bibinfo{person}{Igor Mordatch}, \bibinfo{person}{Zoran Popovi{\'c}}, {and} \bibinfo{person}{Emanuel Todorov}.} \bibinfo{year}{2012}\natexlab{}.
\newblock \showarticletitle{Contact-invariant optimization for hand manipulation}. In \bibinfo{booktitle}{\emph{Proceedings of the ACM SIGGRAPH/Eurographics symposium on computer animation}}. \bibinfo{pages}{137--144}.
\newblock


\bibitem[Peng et~al\mbox{.}(2018)]%
        {deepmimic}
\bibfield{author}{\bibinfo{person}{Xue~Bin Peng}, \bibinfo{person}{Pieter Abbeel}, \bibinfo{person}{Sergey Levine}, {and} \bibinfo{person}{Michiel van~de Panne}.} \bibinfo{year}{2018}\natexlab{}.
\newblock \showarticletitle{DeepMimic: Example-Guided Deep Reinforcement Learning of Physics-Based Character Skills}.
\newblock \bibinfo{journal}{\emph{ACM Trans. Graph.}} \bibinfo{volume}{37}, \bibinfo{number}{4}, Article \bibinfo{articleno}{143} (\bibinfo{year}{2018}).
\newblock


\bibitem[Peng et~al\mbox{.}(2022)]%
        {ase}
\bibfield{author}{\bibinfo{person}{Xue~Bin Peng}, \bibinfo{person}{Yunrong Guo}, \bibinfo{person}{Lina Halper}, \bibinfo{person}{Sergey Levine}, {and} \bibinfo{person}{Sanja Fidler}.} \bibinfo{year}{2022}\natexlab{}.
\newblock \showarticletitle{ASE: Large-Scale Reusable Adversarial Skill Embeddings for Physically Simulated Characters}.
\newblock \bibinfo{journal}{\emph{ACM Trans. Graph.}} \bibinfo{volume}{41}, \bibinfo{number}{4}, Article \bibinfo{articleno}{94} (\bibinfo{year}{2022}).
\newblock


\bibitem[Peng et~al\mbox{.}(2021)]%
        {amp}
\bibfield{author}{\bibinfo{person}{Xue~Bin Peng}, \bibinfo{person}{Ze Ma}, \bibinfo{person}{Pieter Abbeel}, \bibinfo{person}{Sergey Levine}, {and} \bibinfo{person}{Angjoo Kanazawa}.} \bibinfo{year}{2021}\natexlab{}.
\newblock \showarticletitle{AMP: Adversarial Motion Priors for Stylized Physics-Based Character Control}.
\newblock \bibinfo{journal}{\emph{ACM Trans. Graph.}} \bibinfo{volume}{40}, \bibinfo{number}{4}, Article \bibinfo{articleno}{144} (\bibinfo{year}{2021}).
\newblock


\bibitem[Pollard and Zordan(2005)]%
        {pollard2005physically}
\bibfield{author}{\bibinfo{person}{Nancy~S Pollard} {and} \bibinfo{person}{Victor~Brian Zordan}.} \bibinfo{year}{2005}\natexlab{}.
\newblock \showarticletitle{Physically based grasping control from example}. In \bibinfo{booktitle}{\emph{Proceedings of the 2005 ACM SIGGRAPH/Eurographics symposium on Computer animation}}. \bibinfo{pages}{311--318}.
\newblock


\bibitem[Qian et~al\mbox{.}(2021)]%
        {qian2021speech}
\bibfield{author}{\bibinfo{person}{Shenhan Qian}, \bibinfo{person}{Zhi Tu}, \bibinfo{person}{Yihao Zhi}, \bibinfo{person}{Wen Liu}, {and} \bibinfo{person}{Shenghua Gao}.} \bibinfo{year}{2021}\natexlab{}.
\newblock \showarticletitle{Speech drives templates: Co-speech gesture synthesis with learned templates}. In \bibinfo{booktitle}{\emph{Proceedings of the IEEE/CVF International Conference on Computer Vision}}. \bibinfo{pages}{11077--11086}.
\newblock


\bibitem[Radisavljevic and Driessen(2004)]%
        {radisavljevic2004path}
\bibfield{author}{\bibinfo{person}{Aleksander Radisavljevic} {and} \bibinfo{person}{Peter~F Driessen}.} \bibinfo{year}{2004}\natexlab{}.
\newblock \showarticletitle{Path difference learning for guitar fingering problem}. In \bibinfo{booktitle}{\emph{International Conference on Mathematics and Computing}}.
\newblock


\bibitem[Rajeswaran et~al\mbox{.}(2017)]%
        {rajeswaran2017epopt}
\bibfield{author}{\bibinfo{person}{Aravind Rajeswaran}, \bibinfo{person}{Sarvjeet Ghotra}, \bibinfo{person}{Balaraman Ravindran}, {and} \bibinfo{person}{Sergey Levine}.} \bibinfo{year}{2017}\natexlab{}.
\newblock \showarticletitle{{EPO}pt: Learning Robust Neural Network Policies Using Model Ensembles}. In \bibinfo{booktitle}{\emph{Int. Conf. on Learning Representations}}.
\newblock


\bibitem[Schulman et~al\mbox{.}(2017)]%
        {schulman2017proximal}
\bibfield{author}{\bibinfo{person}{John Schulman}, \bibinfo{person}{Filip Wolski}, \bibinfo{person}{Prafulla Dhariwal}, \bibinfo{person}{Alec Radford}, {and} \bibinfo{person}{Oleg Klimov}.} \bibinfo{year}{2017}\natexlab{}.
\newblock \bibinfo{title}{Proximal Policy Optimization Algorithms}.
\newblock
\newblock
\showeprint[arxiv]{1707.06347}~[cs.LG]


\bibitem[Shlizerman et~al\mbox{.}(2018)]%
        {shlizerman2018audio}
\bibfield{author}{\bibinfo{person}{Eli Shlizerman}, \bibinfo{person}{Lucio Dery}, \bibinfo{person}{Hayden Schoen}, {and} \bibinfo{person}{Ira Kemelmacher-Shlizerman}.} \bibinfo{year}{2018}\natexlab{}.
\newblock \showarticletitle{Audio to body dynamics}. In \bibinfo{booktitle}{\emph{Conference on Computer Vision and Pattern Recognition}}. \bibinfo{pages}{7574--7583}.
\newblock


\bibitem[Simon et~al\mbox{.}(2017)]%
        {simon2017hand}
\bibfield{author}{\bibinfo{person}{Tomas Simon}, \bibinfo{person}{Hanbyul Joo}, \bibinfo{person}{Iain Matthews}, {and} \bibinfo{person}{Yaser Sheikh}.} \bibinfo{year}{2017}\natexlab{}.
\newblock \showarticletitle{Hand keypoint detection in single images using multiview bootstrapping}. In \bibinfo{booktitle}{\emph{Conference on Computer Vision and Pattern Recognition}}.
\newblock


\bibitem[Skreinig et~al\mbox{.}(2023)]%
        {skreinig2022guitARhero}
\bibfield{author}{\bibinfo{person}{Lucchas~Ribeiro Skreinig}, \bibinfo{person}{Denis Kalkofen}, \bibinfo{person}{Ana Stanescu}, \bibinfo{person}{Peter Mohr}, \bibinfo{person}{Frank Heyen}, \bibinfo{person}{Shohei Mori}, \bibinfo{person}{Michael Sedlmair}, \bibinfo{person}{Dieter Schmalstieg}, {and} \bibinfo{person}{Alexander Plopski}.} \bibinfo{year}{2023}\natexlab{}.
\newblock \showarticletitle{guitARhero: Interactive Augmented Reality Guitar Tutorials}.
\newblock  \bibinfo{volume}{29}, \bibinfo{number}{11} (\bibinfo{year}{2023}), \bibinfo{pages}{4676--4685}.
\newblock


\bibitem[Taheri et~al\mbox{.}(2020)]%
        {taheri2020grab}
\bibfield{author}{\bibinfo{person}{Omid Taheri}, \bibinfo{person}{Nima Ghorbani}, \bibinfo{person}{Michael~J Black}, {and} \bibinfo{person}{Dimitrios Tzionas}.} \bibinfo{year}{2020}\natexlab{}.
\newblock \showarticletitle{GRAB: A dataset of whole-body human grasping of objects}. In \bibinfo{booktitle}{\emph{Computer Vision--ECCV 2020: 16th European Conference, Glasgow, UK, August 23--28, 2020, Proceedings, Part IV 16}}. Springer, \bibinfo{pages}{581--600}.
\newblock


\bibitem[Wang et~al\mbox{.}(2013)]%
        {wang2013video}
\bibfield{author}{\bibinfo{person}{Yangang Wang}, \bibinfo{person}{Jianyuan Min}, \bibinfo{person}{Jianjie Zhang}, \bibinfo{person}{Yebin Liu}, \bibinfo{person}{Feng Xu}, \bibinfo{person}{Qionghai Dai}, {and} \bibinfo{person}{Jinxiang Chai}.} \bibinfo{year}{2013}\natexlab{}.
\newblock \showarticletitle{Video-based hand manipulation capture through composite motion control}.
\newblock \bibinfo{journal}{\emph{ACM Transactions on Graphics (TOG)}} \bibinfo{volume}{32}, \bibinfo{number}{4} (\bibinfo{year}{2013}), \bibinfo{pages}{1--14}.
\newblock


\bibitem[Won et~al\mbox{.}(2020)]%
        {scadiver}
\bibfield{author}{\bibinfo{person}{Jungdam Won}, \bibinfo{person}{Deepak Gopinath}, {and} \bibinfo{person}{Jessica Hodgins}.} \bibinfo{year}{2020}\natexlab{}.
\newblock \showarticletitle{A scalable approach to control diverse behaviors for physically simulated characters}.
\newblock \bibinfo{journal}{\emph{ACM Transactions on Graphics}} \bibinfo{volume}{39}, \bibinfo{number}{4} (\bibinfo{year}{2020}), \bibinfo{pages}{33--1}.
\newblock


\bibitem[Wortman and Smith(2021)]%
        {wortman2021CombinoChordAG}
\bibfield{author}{\bibinfo{person}{Kevin~A. Wortman} {and} \bibinfo{person}{Nicholas Smith}.} \bibinfo{year}{2021}\natexlab{}.
\newblock \showarticletitle{CombinoChord: A Guitar Chord Generator App}.
\newblock \bibinfo{journal}{\emph{IEEE 11th Annual Computing and Communication Workshop and Conference}} (\bibinfo{year}{2021}), \bibinfo{pages}{0785--0789}.
\newblock


\bibitem[Xie et~al\mbox{.}(2023)]%
        {xie2023hierarchical}
\bibfield{author}{\bibinfo{person}{Zhaoming Xie}, \bibinfo{person}{Jonathan Tseng}, \bibinfo{person}{Sebastian Starke}, \bibinfo{person}{Michiel van~de Panne}, {and} \bibinfo{person}{C~Karen Liu}.} \bibinfo{year}{2023}\natexlab{}.
\newblock \showarticletitle{Hierarchical planning and control for box loco-manipulation}.
\newblock \bibinfo{journal}{\emph{Proceedings of the ACM on Computer Graphics and Interactive Techniques}} \bibinfo{volume}{6}, \bibinfo{number}{3} (\bibinfo{year}{2023}), \bibinfo{pages}{1--18}.
\newblock


\bibitem[Xu and Karamouzas(2021)]%
        {iccgan}
\bibfield{author}{\bibinfo{person}{Pei Xu} {and} \bibinfo{person}{Ioannis Karamouzas}.} \bibinfo{year}{2021}\natexlab{}.
\newblock \showarticletitle{A GAN-Like Approach for Physics-Based Imitation Learning and Interactive Character Control}.
\newblock \bibinfo{journal}{\emph{Proc. of the ACM on Computer Graphics and Interactive Techniques}} \bibinfo{volume}{4}, \bibinfo{number}{3} (\bibinfo{year}{2021}).
\newblock


\bibitem[Xu et~al\mbox{.}(2023a)]%
        {composite}
\bibfield{author}{\bibinfo{person}{Pei Xu}, \bibinfo{person}{Xiumin Shang}, \bibinfo{person}{Victor Zordan}, {and} \bibinfo{person}{Ioannis Karamouzas}.} \bibinfo{year}{2023}\natexlab{a}.
\newblock \showarticletitle{Composite Motion Learning with Task Control}.
\newblock \bibinfo{journal}{\emph{ACM Transactions on Graphics}} \bibinfo{volume}{42}, \bibinfo{number}{4} (\bibinfo{year}{2023}).
\newblock


\bibitem[Xu et~al\mbox{.}(2023b)]%
        {adaptnet}
\bibfield{author}{\bibinfo{person}{Pei Xu}, \bibinfo{person}{Kaixiang Xie}, \bibinfo{person}{Sheldon Andrews}, \bibinfo{person}{Paul~G. Kry}, \bibinfo{person}{Michael Neff}, \bibinfo{person}{Ioannis Karamouzas}, {and} \bibinfo{person}{Victor Zordan}.} \bibinfo{year}{2023}\natexlab{b}.
\newblock \showarticletitle{AdaptNet: Policy Adaptation for Physics-Based Character Control}.
\newblock \bibinfo{journal}{\emph{ACM Transactions on Graphics}} \bibinfo{volume}{42}, \bibinfo{number}{6} (\bibinfo{year}{2023}).
\newblock


\bibitem[Yang et~al\mbox{.}(2022)]%
        {yang2022learning}
\bibfield{author}{\bibinfo{person}{Zeshi Yang}, \bibinfo{person}{Kangkang Yin}, {and} \bibinfo{person}{Libin Liu}.} \bibinfo{year}{2022}\natexlab{}.
\newblock \showarticletitle{Learning to use chopsticks in diverse gripping styles}.
\newblock \bibinfo{journal}{\emph{ACM Transactions on Graphics (TOG)}} \bibinfo{volume}{41}, \bibinfo{number}{4} (\bibinfo{year}{2022}), \bibinfo{pages}{1--17}.
\newblock


\bibitem[Yao et~al\mbox{.}(2022)]%
        {controlvae}
\bibfield{author}{\bibinfo{person}{Heyuan Yao}, \bibinfo{person}{Zhenhua Song}, \bibinfo{person}{Baoquan Chen}, {and} \bibinfo{person}{Libin Liu}.} \bibinfo{year}{2022}\natexlab{}.
\newblock \showarticletitle{ControlVAE: Model-Based Learning of Generative Controllers for Physics-Based Characters}.
\newblock \bibinfo{journal}{\emph{ACM Trans. Graph.}} \bibinfo{volume}{41}, \bibinfo{number}{6}, Article \bibinfo{articleno}{183} (\bibinfo{year}{2022}).
\newblock


\bibitem[Yao et~al\mbox{.}(2023)]%
        {yao2023moconvq}
\bibfield{author}{\bibinfo{person}{Heyuan Yao}, \bibinfo{person}{Zhenhua Song}, \bibinfo{person}{Yuyang Zhou}, \bibinfo{person}{Tenglong Ao}, \bibinfo{person}{Baoquan Chen}, {and} \bibinfo{person}{Libin Liu}.} \bibinfo{year}{2023}\natexlab{}.
\newblock \showarticletitle{MoConVQ: Unified Physics-Based Motion Control via Scalable Discrete Representations}.
\newblock \bibinfo{journal}{\emph{arXiv preprint arXiv:2310.10198}} (\bibinfo{year}{2023}).
\newblock


\bibitem[Ye and Liu(2012)]%
        {ye2012synthesis}
\bibfield{author}{\bibinfo{person}{Yuting Ye} {and} \bibinfo{person}{C~Karen Liu}.} \bibinfo{year}{2012}\natexlab{}.
\newblock \showarticletitle{Synthesis of detailed hand manipulations using contact sampling}.
\newblock \bibinfo{journal}{\emph{ACM Transactions on Graphics (ToG)}} \bibinfo{volume}{31}, \bibinfo{number}{4} (\bibinfo{year}{2012}), \bibinfo{pages}{1--10}.
\newblock


\bibitem[Zakka et~al\mbox{.}(2023)]%
        {zakka2023robopianist}
\bibfield{author}{\bibinfo{person}{Kevin Zakka}, \bibinfo{person}{Philipp Wu}, \bibinfo{person}{Laura Smith}, \bibinfo{person}{Nimrod Gileadi}, \bibinfo{person}{Taylor Howell}, \bibinfo{person}{Xue~Bin Peng}, \bibinfo{person}{Sumeet Singh}, \bibinfo{person}{Yuval Tassa}, \bibinfo{person}{Pete Florence}, \bibinfo{person}{Andy Zeng}, {et~al\mbox{.}}} \bibinfo{year}{2023}\natexlab{}.
\newblock \showarticletitle{Robopianist: Dexterous piano playing with deep reinforcement learning}. In \bibinfo{booktitle}{\emph{Conference on Robot Learning}}.
\newblock


\bibitem[Zhang et~al\mbox{.}(2021)]%
        {zhang2021manipnet}
\bibfield{author}{\bibinfo{person}{H Zhang}, \bibinfo{person}{Y Ye}, \bibinfo{person}{T Shiratori}, {and} \bibinfo{person}{T Komura}.} \bibinfo{year}{2021}\natexlab{}.
\newblock \showarticletitle{ManipNet: neural manipulation synthesis with a hand-object spatial representation}.
\newblock \bibinfo{journal}{\emph{ACM Transactions on Graphics}} (\bibinfo{year}{2021}).
\newblock


\bibitem[Zhang et~al\mbox{.}(2024)]%
        {Zhang2024SemanticGesture}
\bibfield{author}{\bibinfo{person}{Zeyi Zhang}, \bibinfo{person}{Tenglong Ao}, \bibinfo{person}{Yuyao Zhang}, \bibinfo{person}{Qingzhe Gao}, \bibinfo{person}{Chuan Lin}, \bibinfo{person}{Baoquan Chen}, {and} \bibinfo{person}{Libin Liu}.} \bibinfo{year}{2024}\natexlab{}.
\newblock \showarticletitle{Semantic Gesticulator: Semantics-Aware Co-Speech Gesture Synthesis}.
\newblock \bibinfo{journal}{\emph{ACM Trans. Graph.}} (\bibinfo{year}{2024}), \bibinfo{numpages}{17}~pages.
\newblock
\urldef\tempurl%
\url{https://doi.org/10.1145/3658134}
\showDOI{\tempurl}


\bibitem[Zhao et~al\mbox{.}(2013)]%
        {zhao2013robust}
\bibfield{author}{\bibinfo{person}{Wenping Zhao}, \bibinfo{person}{Jianjie Zhang}, \bibinfo{person}{Jianyuan Min}, {and} \bibinfo{person}{Jinxiang Chai}.} \bibinfo{year}{2013}\natexlab{}.
\newblock \showarticletitle{Robust realtime physics-based motion control for human grasping}.
\newblock \bibinfo{journal}{\emph{ACM Transactions on Graphics (TOG)}} \bibinfo{volume}{32}, \bibinfo{number}{6} (\bibinfo{year}{2013}), \bibinfo{pages}{1--12}.
\newblock


\bibitem[Zhu et~al\mbox{.}(2013)]%
        {zhu2013piano}
\bibfield{author}{\bibinfo{person}{Yuanfeng Zhu}, \bibinfo{person}{Ajay~Sundar Ramakrishnan}, \bibinfo{person}{Bernd Hamann}, {and} \bibinfo{person}{Michael Neff}.} \bibinfo{year}{2013}\natexlab{}.
\newblock \showarticletitle{A system for automatic animation of piano performances}.
\newblock \bibinfo{journal}{\emph{Computer Animation and Virtual Worlds}} \bibinfo{volume}{24}, \bibinfo{number}{5} (\bibinfo{year}{2013}), \bibinfo{pages}{445--457}.
\newblock


\end{thebibliography}

\appendix

\renewcommand{\thefigure}{S\arabic{figure}}
\renewcommand{\thetable}{S\arabic{table}}
\def\theequation{S\arabic{equation}}
\setcounter{figure}{0}
\setcounter{table}{0}
\setcounter{equation}{0}
\setcounter{page}{1}

\section{Guitar Playing Motion Dataset}
\begin{figure}[t]
    \centering
    \includegraphics[width=\linewidth]{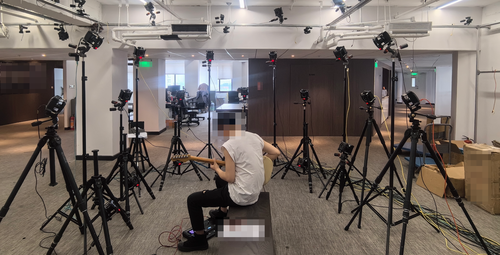}\vspace{0.3em}\\

    \includegraphics[width=.49\linewidth]{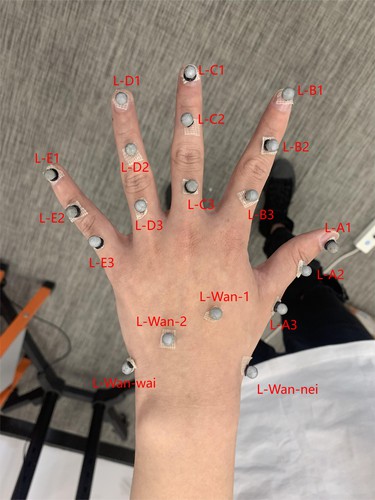}\hfill\includegraphics[width=.49\linewidth]{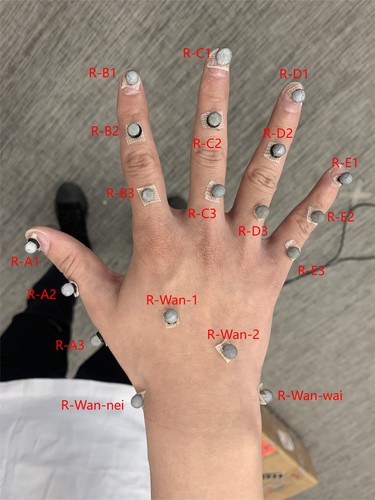}\vspace{0.3em}\\
    
    \includegraphics[width=\linewidth]{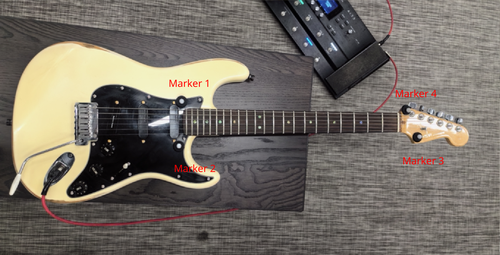}
    \caption{Setup of our motion capture pipeline. Top: camera deployment. Middle: marker placement on hands. Bottom: maker placement on guitar.}
    \label{fig:mocap}
\end{figure}
Our motion data were captured from a professional guitarist.
We took the motion capture solutions provided by NOKOV and employed 18 infrared cameras for motion capture, 7 of which have a resolution of 9MP and 11 have a resolution of 4MP.
We placed 19 markers on each hand of the subject, and 4 markers on the guitar for localization.
All data were captured at 120FPS.
The camera deployment and marker placement are exhibited in Fig.~\ref{fig:mocap}.
We collected diverse general guitar-playing motions about 1 hour long, including:
\begin{itemize}
    \item 12 major scales,
    \item chromatic scales,
    \item diverse chords,
    \item arpeggios,
    \item strumming and picking,
    \item bends,
    \item sliding,
    \item vibrato,
    \item palm mute,
    \item natural harmonics,
    \item artificial harmonics,
    \item hammer-ons and pull-offs.
\end{itemize}
All notes were played as 1/8th and mostly at 100BPM. 
The instrument is a \textit{Fender American Deluxe Stratocaster}\footnote{https://serialnumberlookup.fender.com/product/0101000748} guitar with 6 strings and 22 frets. The scale length is 25.5 inches.

\section{Implementation Details}

\begin{table}[t]
\centering
\caption{Hyperparameters}
\begin{tabular}{lc}
    \toprule
    \textbf{Parameter} & \textbf{Value}\\
    \midrule
    policy network learning rate & $5 \times 10^{-6}$\\
    critic network learning rate & $1 \times 10^{-4}$\\
    discriminator learning rate & $1 \times 10^{-5}$\\
    reward discount factor ($\gamma$) & $0.95$ \\
    GAE discount factor ($\lambda$) & $0.95$ \\
    surrogate clip range ($\epsilon$) & $0.2$ \\
    gradient penalty coefficient ($\lambda^{GP}$) & $10$ \\
    number of PPO workers (simulation instances) & $512$ \\
    PPO replay buffer size & $4096$ \\
    PPO batch size & $256$ \\
    PPO optimization epochs & $5$ \\
    discriminator replay buffer size & $8192$ \\
    discriminator batch size & $512$ \\
  \bottomrule
\end{tabular}
\label{tab:hyper}
\end{table}

\begin{figure}[t]
    \centering
    \begin{subfigure}[t]{.3\linewidth}
    \includegraphics[width=\linewidth]{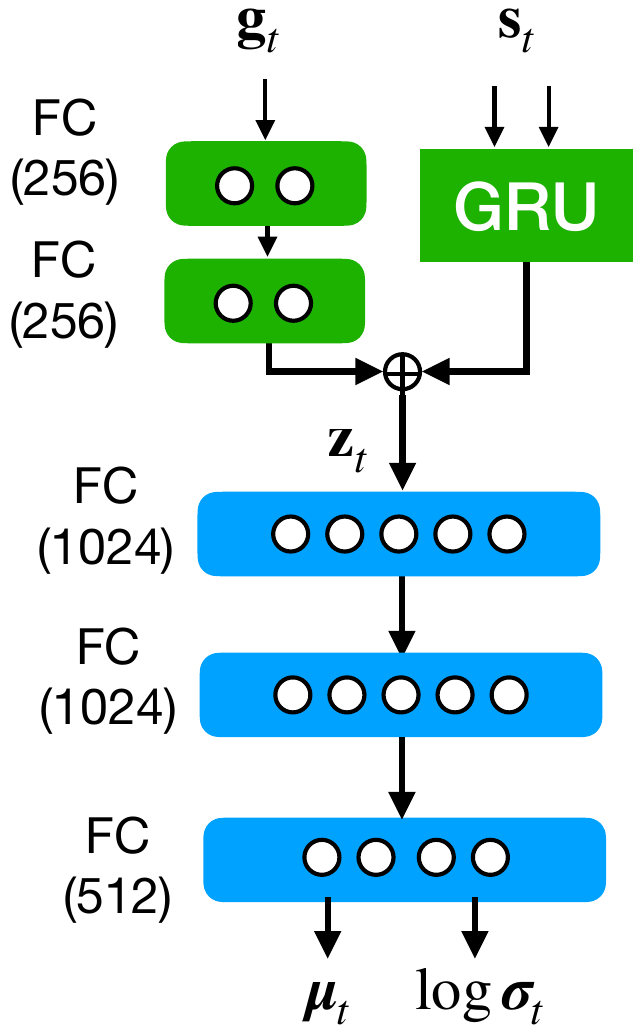}
    \caption{Policy Network}
    \end{subfigure}\quad
    \begin{subfigure}[t]{.3\linewidth}\centering
    \includegraphics[width=\linewidth]{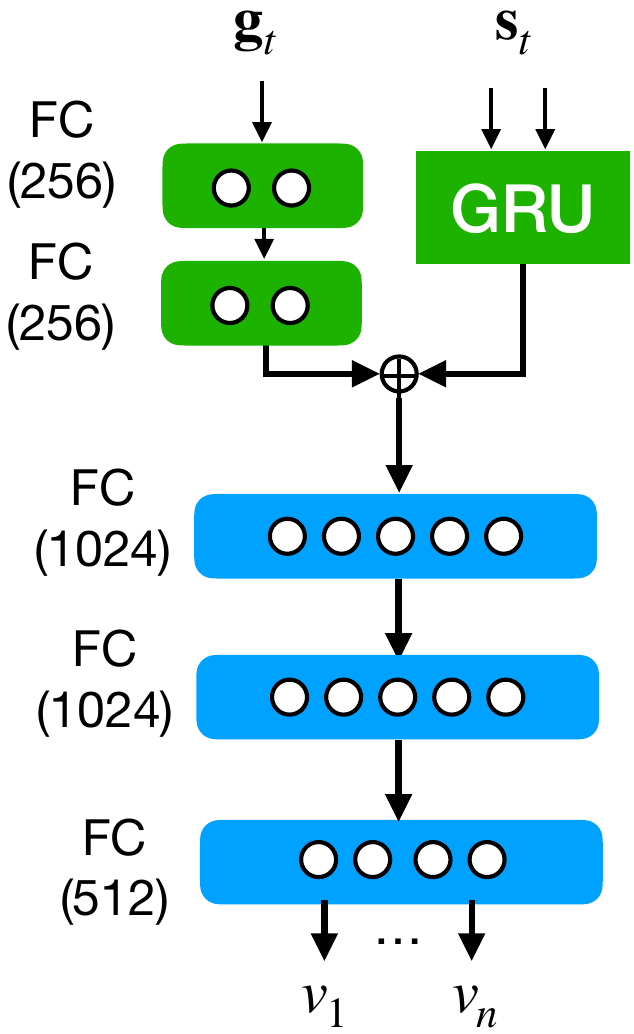}
    \caption{Value Network}
    \end{subfigure}
    \begin{subfigure}[t]{.35\linewidth}\centering
    \includegraphics[width=.81\linewidth]{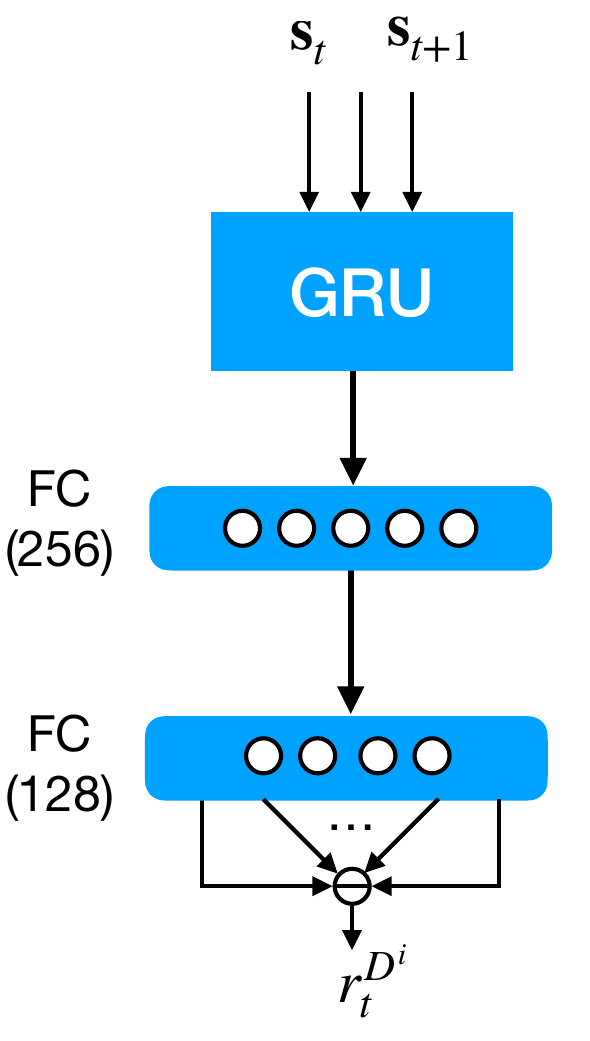}
    \caption{Discriminator Network}
    \end{subfigure}
    \caption{Network structures. We use $\oplus$ denoting the add operator and $\ominus$ denoting the average operator. The state encoder consists of blocks in green, where a GRU encoder is employed for pose state encoding temporally, and a two-layer MLP encoder is employed for goal state encoding.}
    \label{fig:network}
\end{figure}

All our control policies are optimized using PPO~\cite{schulman2017proximal} as the backbone reinforcement learning algorithm with network parameters updated by Adam optimizers~\cite{kingma2014adam}.
The hyperparameters used for policy training are listed in Table~\ref{tab:hyper}.
Half of the training samples used by the discriminator are drawn
from the simulated character and half are sampled from the reference
motions.
The network structures are shown in Fig.~\ref{fig:network}.
We use a GAN-like architecture combining reinforcement learning and motion imitation for policy training~\cite{iccgan}.
We use the same network structure for the left and right-hand policies, while our policy synchronization scheme does not require them to be identical.
The state encoder of each policy consists of a GRU~\cite{chung2014empirical} encoder with a 256-dimensional hidden state for pose state encoding and an MLP encoder for goal state encoding. 
The two encoded states are added together as the latent for manipulation during policy synchronization.
The value networks have the same architecture as the policy networks, but have multiple outputs depending on the number of objectives (8 for the left-hand policy, 2 for the right-hand policy, and 9 for the joint policy synchronized using pre-trained single-hand policies).
Especially, when training two-hand policies,
it takes $\mathbf{s}_t = \{\mathbf{s}_t^L, \mathbf{s}_t^R\}$ and $\mathbf{g}_t = \{\mathbf{g}_t^L, \mathbf{g}_t^{R+}\}$ as input.
We normalize the fretting representation of $\mathbf{g}_t^L$ into the range of $[-1, 1]$ for the left-hand policy, while the right-hand policy uses a binary representation of $\mathbf{g}_t^R$ to indicate if a string should be picked or not.
We run a moving-average normalizer on $\mathbf{s}_t^L$ and $\mathbf{s}_t^R$ during policy training. For value networks, we perform output normalization using the technique of PopArt~\cite{rajeswaran2017epopt}.

All policies were trained on a machine equipped with a Nvidia V100 GPU.
It takes around 3-4 days to perform individual policy training, consuming about $8\times10^8$ samples.
The synchronization can be done quickly, typically within 1.5 to 12 hours consuming about $1\times10^7$ to $8\times10^7$ samples, depending on the difficulty of the songs (mainly related to the chord patterns and tempos).

\section{Ablation Studies}
\begin{figure}
    \centering
    \includegraphics[width=\linewidth]{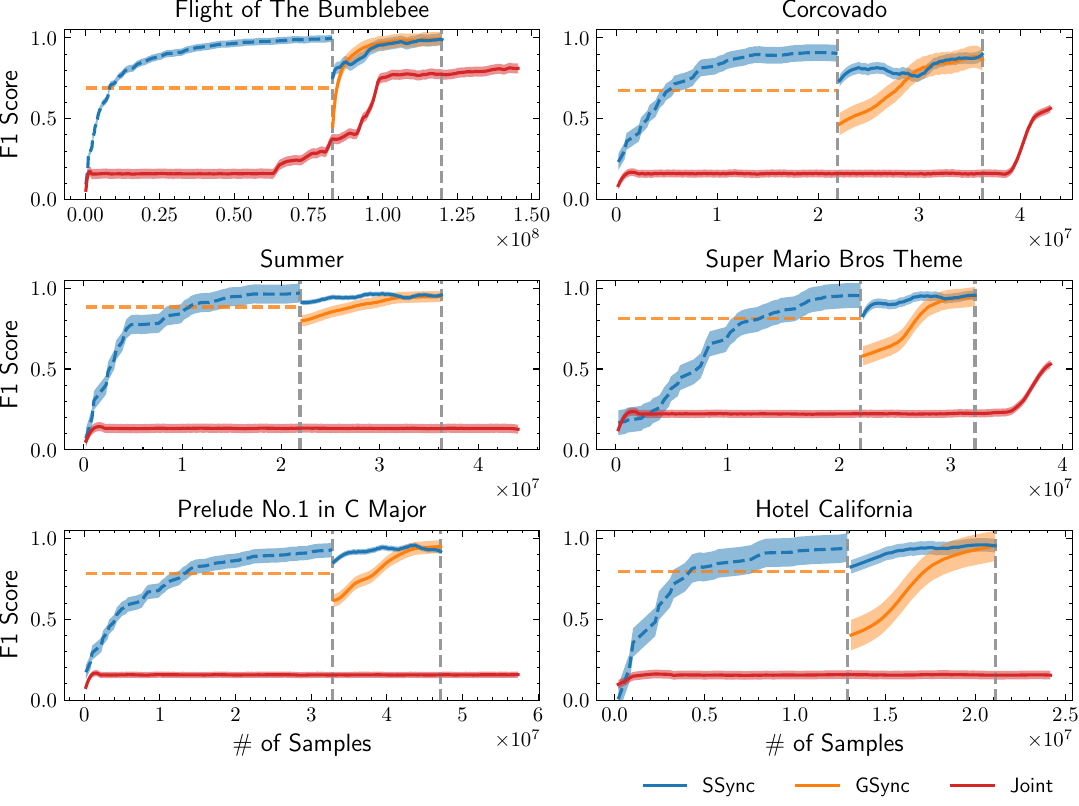}
    \caption{Training performance of the two-hand policies using the pre-trained general-purpose left-hand policy for synchronization (\textit{GSync}) and that using policies trained only for specific songs (\textit{SSync}). The dashed orange and blue lines indicate the corresponding left-hand policies' performance when evaluated solely without the right hand. For the pre-trained general-purpose policy, we show its evaluation performance using dashed orange lines. For the specifically trained policies, we show their training performance using dashed blue lines. The gray lines are an indicator to distinguish the single-hand policy training phase and the synchronization phase.
    Red lines are the learning performance when two hands are put together as a joint agent trained from scratch.}
    \label{fig:ablation}
\end{figure}
Our proposed policy synchronization approach can synchronize the behaviors of two hands for cooperative bimanual control, and, meanwhile, fine-tune the pre-trained single-hand policies for specific songs.
Both of them would bring about an improvement in the F1 scores of the achieved two-hand policies.
To show the performance of policy synchronization solely, 
we study the synchronization performance using left-hand policies that are trained specially for specific songs.
We still use the right-hand policy that was pre-trained for general purposes,
as it already can perform string picking accurately for all the tested songs.

In Fig.~\ref{fig:ablation},
we compare the two-hand performance during training using the pre-trained general-purpose left-hand policy (\textit{GSync}) and that using the ones trained for specific songs (\textit{SSync}).
The performances of the corresponding single-hand policies are shown in dashed lines.
As can be seen,
typically,  training a left-hand policy for a specific song (dashed lines in blue) would consume around $2\times10^7$ samples.
For difficult pieces, e.g. the very fast song \textit{Flight of The Bumblebee} that has a tempo of 170BPM and the classic music \textit{Prelude No.1 in C Major} that has lots of arpeggios,
it would need around $3.5\times10^7$ and even $8\times10^7$ samples.
The specially trained policies can reach an F1 score above 0.98 or even equal to 1, which is nearly perfect in terms of string-pressing accuracy.
However, when put together with the right hand for cooperation,
they still suffer a significant performance drop of around 20\%,
because of the desynchronization between the two hands' behaviors.
The performance of the pre-trained general-purpose left-hand policy (dashed lines in orange)
is worse than the specially trained left-hand policies, especially for songs with many chords, e.g. \textit{Corcovado}, but still can provide an F1 score above 0.8 for most of the tested songs.
While being able to reach a similar final performance with the specially trained left-hand policies through synchronization, 
the general-purpose policy is reusable for different songs.
We can also see that the worse performance before policy synchronization, as shown in Fig.~\ref{fig:f1_sync} mainly comes from the unsynchronized behaviors of the single-hand policies. 
Our policy synchronization scheme can effectively coordinate the behaviors of two hands within a small budget of training samples, and achieve high F1 scores at the end.

Additionally, in Fig.~\ref{fig:ablation},
we show the training curves when two hands are directly trained together from scratch as a joint agent (\textit{Joint}).
In that case, we employ the same network structure as that used for single-hand policy training, but use a concatenated pose state of $\mathbf{s}_t^L$ and $\mathbf{s}_t^R$. The policy is optimized under the same training scheme depicted in Section~\ref{sec:guitar_sync} but has one more imitation objective for the right hand.
As shown in the figure, 
training a two-hand policy directly from scratch is difficult.
Though we can see a performance improvement for some tested songs as the training goes on, 
the curve increases very slowly.
The two hands, at the initial training phase, mainly perform random exploration for basic skill learning rather than trying to build effective cooperation,
and thus the learning is inefficient within the limited budget of training samples.
Note that simulating scenarios with two hands is around 40\% slower than that with only one hand.
This means that when the same number of samples are consumed,
it would take much more time to train a two-hand policy.
Therefore, even if the \textit{Joint} policy may finally achieve better performance as the training goes on, 
our proposed approach is more efficient by reducing the training needed for the joint policy training of two hands in the centralized environment.
We refer to the supplementary video for visual comparison.

\section{Limitations}
For the left hand, policies trained by our approach can perform fret pressing in natural human poses with high accuracy.
The resulting motions, though dexterous, are not always able to fully reflect the grace of human guitarists in terms of smooth hand movement.
Ideally, accurate fretting requires swift movement of the hand and fingers to press target notes responsively, minimizing the transitioning time between consecutive notes as much as possible.
This, however, will result in fast and abrupt movement.
While professional human guitarists would exhibit more sophisticated planning behaviors for energy-efficient motion control,
our energy-related reward term (Eq.~\ref{eq:r_energy_L}) simply penalizes the hand and fingers' fast movement 
at the cost of responsiveness.
The associated weight for the term has to be small enough to guarantee the policy's responsiveness even when facing notes lasting for short durations (cf. Eq.~\ref{eq:r_il} and~\ref{eq:r_ir}, or has to be fine-tuned carefully for specific rhythms to balance responsiveness and energy efficiency. 
Due to this trade-off,
our control policies cannot completely avoid unnatural fast or abrupt movement of the hand and fingers, especially for fast songs,
as we can see from the animated results shown in the supplementary video.
Besides, our policies only have a limited horizon of future notes (5 in our implementation), while experienced human guitarists would perform more careful and intelligent motion planning for fretting based on a longer horizon and other factors like finger dexterity and muscle fatigue.
In the arpeggio demo shown in Fig.~\ref{fig:arpeggios},
due to the limited observation horizon, the left-hand policy itself is unable to detect chord patterns based on the target notes of individual strings.
We have to specially feed a chord note as the goal input for the left hand while the right hand takes the normal goal notes of individual strings, or the left hand will press single strings without keeping a consistent chord-pressing pose. 
Though this is analogous to explicitly putting a chord symbol on the tab to guide fretting in real-world practice,
it would be an interesting direction for future work to further introduce human-level intelligence for motion planning during guitar playing.

Another limitation of this work is the lack of physically simulated string dynamics.
This makes us have to rely on heuristic rules to decide the pressed and picked states of strings and generate sound by post-processing,
rather than directly getting their physical vibration states for accurate sound generation.
The lack of string dynamics constrains us from introducing guitar-playing techniques that involve complex interactions with strings, like vibrato and fingerstyle.
Meanwhile, we use rigid bodies instead of deformable ones for hand modeling, which may overlook subtle changes in hand shape.
Besides, all simulations run discretely with a fixed FPS (60Hz in our implementation), limiting the control's time resolution.
For example, given music where the shortest notes are 1/16th notes,
tempos between 100 and 105BPM will be rounded as 100 to ensure that the shortest notes last for an integer number of frames.
It is a
great direction for future work to address the fine control problem
with arbitrary temporal resolution requirements.

As a general approach for two-agent cooperative learning,
our study in the guitar-playing task demonstrates that
the proposed policy synchronization scheme can efficiently adapt two single-agent policies into a joint policy for cooperative tasks.
However, this scheme requires decentralized environments for individual policy training, which are not always easy to set up.
As a bimanual control method for instrument playing,
our approach is expected to work for instruments similar to guitar, like bass and ukulele, but may not be generalized to those involving complex string dynamics like harp and violin. Developing unified policies for various instruments is an interesting direction for future research.

\end{document}